\documentclass[12pt,a4paper]{article}
\usepackage[utf8]{inputenc}
\usepackage{a4wide}
\usepackage{graphicx}
\usepackage{axodraw}
\usepackage{hyperref}
\newcommand{\be}{\begin{equation}}
\newcommand{\ee}{\end{equation}}
\newcommand{\ba}{\begin{eqnarray}}
\newcommand{\ea}{\end{eqnarray}}
\begin{document}

\begin{titlepage}
\begin{flushright}
LU TP 14-38\\
arXiv:1411.6384 [hep-lat]\\
Revised January 2015
\end{flushright}
\vfill
\begin{center}
{\Large\bf Finite Volume at Two-loops\\[4mm]in Chiral Perturbation Theory}
\vfill
{\bf Johan Bijnens and Thomas Rössler}\\[0.3cm]
{Department of Astronomy and Theoretical Physics, Lund University,\\
S\"olvegatan 14A, SE 223-62 Lund, Sweden}
\end{center}
\vfill
\begin{abstract}
We calculate the finite volume corrections to meson masses and decay constants
in two and three flavour Chiral Perturbation Theory to two-loop order.
The analytical results are compared with the existing result for the pion
mass in two-flavour ChPT and the partial results for the other quantities.
We present numerical results for all quantities.
\end{abstract}
\vfill
\vfill
\end{titlepage}


\section{Introduction}

Lattice QCD now provides good calculations of a number of quantities
relevant for low-energy particle physics as reviewed in \cite{Aoki:2013ldr}.
These need several extrapolations, in the quark masses, in the lattice
spacing, in the lattice size and in lattice artefacts. 
Chiral Perturbation Theory (ChPT)
\cite{Weinberg:1978kz,Gasser:1983yg,Gasser:1984gg}
provides guidance for all of these extrapolations. In particular, it can be
used to estimate the corrections due to the finite lattice size.
This was introduced by Gasser and Leutwyler in
\cite{Gasser:1986vb,Gasser:1987ah,Gasser:1987zq}.
This is an alternative method
compared to the one introduced by L\"uscher \cite{Luscher:1985dn} where the
leading finite size corrections can be derived using the scattering amplitude.

In this paper we will restrict ourselves to the $p$-regime with
$m_\pi L >> 1$. We will not do the all order integration over the zero mode
as is necessary in the so-called
$\epsilon$-regime \cite{Gasser:1986vb,Gasser:1987ah,Gasser:1987zq}.
The finite volume corrections to the mass and decay constant
in the equal mass case to one-loop order were calculated in these original
papers. Since then, there have been many studies of finite size effects at
one-loop order in ChPT, in particular the masses and decay constants
to that order were derived in \cite{Becirevic:2003wk} and
\cite{DescotesGenon:2004iu}.

In infinite volume the ChPT expressions for masses and decay
constants are known for all relevant cases and including a number of extensions
as e.g. partially quenched ChPT to two-loop order. This is reviewed
in \cite{Bijnens:2006zp}. There exist a few two-loop calculations at
finite volume in ChPT. The mass in two-flavour ChPT was studied
in \cite{Colangelo:2006mp} and the
quark-anti quark vacuum expectation value in three-flavour ChPT in
\cite{Bijnens:2006ve}, the latter can be extended to the $\epsilon$-regime
\cite{Damgaard:2008zs}. 

The main purpose of this paper is to provide the two-loop finite volume
expressions in two and three-flavour ChPT for the masses and decay constants.
The extension to partially quenched ChPT is planned for future work.
The main reason this was not done earlier is the complexity of the
sunset integral at finite volume. The needed integrals have been recently
worked out in \cite{Bijnens:2013doa}. We will use their expressions
extensively. Our expressions are valid in the frame with $\vec p=0$, often
called the center-of-mass frame. In the so-called moving frames
or with twisted boundary conditions there will be additional terms.

Some preliminary numerical results were reported in \cite{talk}.
We find the typical $e^{-m_\pi L}$ behaviour for most quantities as
expected. The corrections for the pion mass and decay constant are significant
at the present lattice size and precision in lattice QCD calculations.
The corrections for the kaon decay constant are needed but are not quite as
large. The kaon mass has corrections below 1\% and the corrections
for the eta mass and decay constant turn out to be negligible at present
precision. These results are in qualitative agreement with the earlier
work.

We give a short list of references for ChPT and discuss some small points in
Sect.~\ref{ChPT}. The definitions of the integrals we use and how they
relate to the results in \cite{Bijnens:2013doa} is given in
Sect.~\ref{finitevolumeintegrals}. The next section contains our first major
results. The full finite volume correction to the pion mass and decay
constant to two-loop order in ChPT. Sect.~\ref{sec:nf3} contains the
results for the three-flavour case for pion, kaon and eta for both the
mass and decay constant but the large two-loop order formulas are
collected in the appendices. The detailed numerical discussion of our results
is in Sect.~\ref{numerics}.

\section{Chiral Perturbation Theory}
\label{ChPT}

An introduction to ChPT can be found in \cite{Scherer} and in the
two-loop review \cite{Bijnens:2006zp}. The lowest order and $p^4$-Lagrangian
can be found in \cite{Gasser:1983yg} and \cite{Gasser:1984gg} for the
two and three flavour case respectively. The order $p^6$ Lagrangian is given
in \cite{Bijnens:1999sh}. We use the standard renormalization scheme in ChPT.
The needed part for the finite volume integrals is discussed in 
Sect.~\ref{finitevolumeintegrals}. An extensive discussion of the scheme can
be found in \cite{Bijnens:1997vq} and \cite{Bijnens:1999hw}.
An important comment is that the LECs do not depend on the volume \cite{Gasser:1987zq}.

We prefer to designate orders by the $p$-counting order at which the diagram
appears. Thus we refer to order $p^2$, order $p^4$ or one-loop order
and order $p^6$ or two-loop order and include in the terminology
one- or two-loop order also the diagrams with fewer loops but the same order in $p$-counting.

We present the formulas here in terms of the physical \emph{infinite} volume
masses and decay constants.

\section{Comments on the finite volume integrals}
\label{finitevolumeintegrals}

The loop integrals at finite volume at one-loop are well known.
The difference with infinite volume is that there is a sum over discrete
momenta in every direction with a finite size rather than a continuous integral.
The use of the Poisson summation formula allows to identify the
infinite volume part and the finite volume corrections.
The remainder can be done in two ways. For one-loop tadpole
integrals the first one was introduced in the original
work \cite{Gasser:1986vb,Gasser:1987ah,Gasser:1987zq} and one remains
with a sum over Bessel functions, that for large $ML$ converges fast.
The other method can be found in \cite{Becirevic:2003wk} and one remains with
an integral over a Jacobi theta function, this method can be used for small
and medium $ML$ as well. The extensions to other one-loop
integrals can be done in both cases by combining propagators with Feynman
parameters. The first method was extended to the equal mass two-loop
sunset integral in \cite{Colangelo:2006mp}. The general mass case was then done
in both methods in \cite{Bijnens:2013doa}. The methods are explained in detail
in \cite{Bijnens:2013doa} for both the one and two-loop case.
Note that here we use Minkowski notation for the integrals.

The tadpole integrals $A$ and $A_{\mu\nu}$ are defined via
\be
\left\{A(m^2),A_{\mu\nu}(m^2)\right\}
 = \frac{1}{i}\int_V\frac{d^d r}{(2\pi)^d}
\frac{\left\{1,r_\mu r_\nu\right\}}{(r^2-m^2)}\,.
\ee
The $B^0$ tadpole integrals are defined similarly with a doubled propagator,
alternatively as the derivative w.r.t. $m^2$ of the $A$-tadpoles.
The subscript $V$ on the integral indicates that the integral is a discrete sum
over the three spatial components and an integral over the remainder. 
At finite volume, there are more Lorentz-structures possible. 
We define the tensor $t_{\mu\nu}$ as the spatial part of the Minkowski metric
$g_{\mu\nu}$, to express these. For the center-of-mass (cms) case this is sufficient.
The needed functions for $A_{\mu\nu}$ are
\be
A_{\mu\nu}(m^2) = g_{\mu\nu}A_{22}(m^2)+t_{\mu\nu}A_{23}(m^2)\,.
\ee
In infinite volume $A_{22}$ can be rewritten in terms of $A$. At finite volume,
the relation is
\be
dA_{22}(m^2)+3A_{23}(m^2) = m^2 A(m^2)\,.
\ee
This is used to remove $A_{22}$ from our expressions. In addition we do an
expansion in $\epsilon$ with $d=4-2\epsilon$ via
\be
A(m^2) = \lambda_0 \frac{m^2}{16\pi^2}+\overline A(m^2)+A^V(m^2)
+\epsilon\left(A^\epsilon(m^2)+A^{V\epsilon}(m^2)\right)+\cdots\,.
\ee
with $\lambda_0=\frac{1}{\epsilon}+\log(4\pi)+1-\gamma$ and similarly for
the other one-loop integrals. $\lambda_0$ corresponds to the
usual $\overline{MS}$ variant used in ChPT. Doing the renormalization
introduces a subtraction point dependence which corresponds to using for
$\overline{A}(m^2)$ and $\overline{B}^0(m^2)$
\be
\overline{A}(m^2) = \frac{-m^2}{16\pi^2}\log\frac{m^2}{\mu^2}\,,
\quad
\overline{B}^0(m^2) = \frac{-1}{16\pi^2}\left(\log\frac{m^2}{\mu^2}+1\right)\,.
\ee

The sunset integrals are defined as
\ba
\label{defH}
\lefteqn{\left\{H,H_\mu,H_\mu^s,H_{\mu\nu},H_{\mu\nu}^{rs},H_{\mu\nu}^{ss}\right\}
(m_1^2,m_2^2,m_3^2,p) =}
&&\nonumber\\&&
\frac{1}{i^2}\int_V\frac{d^d r}{(2\pi)^d}\frac{d^d s}{(2\pi)^d}
\frac{\left\{1,r_\mu,s_\mu,r_\mu r_\nu, r_\mu s_\nu, s_\mu s_\nu\right\}}
{\left(r^2-m_1^2\right)\left(s^2-m_2^2\right)\left((r+s-p)^2-m_3^2\right)}\,.
\ea
The subscript $V$ again indicates that the spatial dimensions are a discrete
sum rather than an integral.
The conventions correspond to those in infinite volume of
\cite{Amoros:1999dp}. The interchange $r,m_1^2\leftrightarrow s,m_2^2$
shows that $H^s_\mu, H^{ss}_{\mu\nu}$ are related directly to
$H^r_\mu,H^{rr}_{\mu\nu}$. $H^{rs}_{\mu\nu}$ can also be related to
$H_{\mu\nu}$ using the trick shown in \cite{Amoros:1999dp} which remains
valid at finite volume in the cms frame \cite{Bijnens:2013doa}.

In the cms frame we define the functions\footnote{In the cms frame
we have that $t_{\mu\nu}= g_{\mu\nu}-p_\mu p_\nu/p^2$ but the given separation
appears naturally in the calculation \cite{Bijnens:2013doa}. It also avoids
singularities in the limit $p\to0$.}
\ba
\label{defHi}
H_\mu &=& p_\mu H_1\,
\\\nonumber
H_{\mu\nu} &=& p_\mu p_\nu H_{21} + g_{\mu\nu} H_{22} + t_{\mu\nu} H_{27}\,.
\ea
The arguments of all functions in the cms frame are
$(m_1^2,m_2^2,m_3^2,p^2)$.
These functions satisfy the relations, valid in finite
volume \cite{Bijnens:2013doa},
\ba
\label{Hrelations}
H_1(m_1^2,m_2^2,m_3^2,p^2)+
H_1(m_2^2,m_3^2,m_1^2,p^2)+
H_1(m_3^2,m_1^2,m_2^2,p^2)
&=& H(m_1^2,m_2^2,m_3^2,p^2)\,,
\nonumber\\
p^2 H_{21}+d H_{22} + 3 H_{27}-m_1^2 H &=& A(m_2^2)A(m_3^2)\,.
\ea
The arguments of the sunset functions in the second relation are
all $(m_1^2,m_2^2,m_3^2,p^2)$.
These relations have been used to remove $H_{22}$ from the final result
and simplify the expressions somewhat.

We now split the functions in an infinite volume part $\tilde H_i$ and a finite
volume correction $\tilde H^V_i$ with
$H_i=\tilde H_i+\tilde H^V_i$. The infinite volume part was
derived in \cite{Amoros:1999dp}. For the finite volume parts we define
\ba
\tilde H^V &=& \frac{\lambda_0}{16\pi^2}
 \left(A^V(m_1^2)+A^V(m_2^2)+A^V(m_3^2)\right)
 +\frac{1}{16\pi^2}
    \left(A^{V\epsilon}(m_1^2)+A^{V\epsilon}(m_2^2)+A^{V\epsilon}(m_3^2)\right)
\nonumber\\&&
 +H^V\,,
\nonumber\\
\tilde H^V_1 &=& \frac{\lambda_0}{16\pi^2}\frac{1}{2}
 \left(A^V(m_2^2)+A^V(m_3^2)\right)
 +\frac{1}{16\pi^2}\frac{1}{2}
    \left(A^{V\epsilon}(m_2^2)+A^{V\epsilon}(m_3^2)\right)
 +H^V_1\,,
\nonumber\\
\tilde H^V_{21} &=& \frac{\lambda_0}{16\pi^2}\frac{1}{3}
 \left(A^V(m_2^2)+A^V(m_3^2)\right)
 +\frac{1}{16\pi^2}\frac{1}{3}
    \left(A^{V\epsilon}(m_2^2)+A^{V\epsilon}(m_3^2)\right)
 +H^V_{21}\,,
\nonumber\\
\tilde H^V_{27} &=& \frac{\lambda_0}{16\pi^2}
 \left(A^V_{23}(m_1^2)+\frac{1}{3}A_{23}(m_2^2)+\frac{1}{3}A^V_{23}(m_3^2)\right)
\nonumber\\&&
 +\frac{1}{16\pi^2}
    \left(A^{V\epsilon}_{23}(m_1^2)
+\frac{1}{3}A^{V\epsilon}_{23}(m_2^2)+\frac{1}{3}A^{V\epsilon}_{23}(m_3^2)\right)
 +H^V_{27}\,,
\ea
Note that the finite parts are defined slightly different compared to the
infinite volume definition in \cite{Amoros:1999dp}. Here we have pulled out the
extra parts with $A^{V\epsilon}$. These functions cancel in the final
result. We will also use the derivatives w.r.t. $p^2$ of the sunset integrals.
These we denote with and extra prime,
$H^{V\prime}_i\equiv(\partial/\partial p^2)H^V_i$.

The functions $H^V_i$ can be computed with the methods
of \cite{Bijnens:2013doa}.
They correspond to adding the parts labeled with $G$ and $H$ in Sect. 4.3 and
the part of Sect. 4.4 in \cite{Bijnens:2013doa}.
We have in addition added the derivatives w.r.t. $p^2$ for all the integrals
and checked the analytical results with numerical differentiation.

For all cases discussed we have done checks that both methods, via Bessel or
Jacobi theta functions, give the same results.
 
\section{Two-flavour results}
\label{sec:nf2}

\begin{figure}[tb!]
\begin{center}
\includegraphics[width=0.6\textwidth]{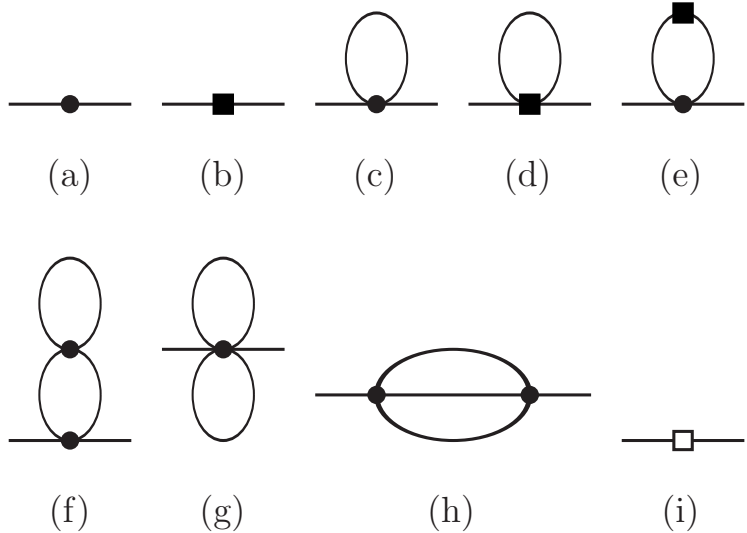}
\end{center}
\caption{The Feynman diagrams needed for the mass calculation. A dot indicates a vertex of order $p^2$, a filled box of order $p^4$ and an open box of order $p^6$.}
\label{figdiagrams}
\end{figure}

The diagrams needed to obtain the mass are shown in Fig.~\ref{figdiagrams}.
We write the result for the mass at finite volume in the form
\be
m_\pi^{V2} = m_\pi^2+\Delta^V\! m^2_\pi\,,
\quad \Delta^V\! m^2_\pi = \Delta^V\! m^{2(4)}_\pi+\Delta^V\! m^{2(6)}_\pi\,.
\ee
$m_\pi^2$ and $F_\pi$ denote the infinite volume physical pion mass and decay
constant.
We have reproduced the expression for the infinite volume mass derived
in \cite{Burgi:1996qi,Bijnens:1995yn,Bijnens:1998fm}.
The extra parts due to the finite volume are
\ba
F_\pi^2 \Delta^V\! m^{2(4)}_\pi&=&
 -\frac{1}{2}m_\pi^2 {A}^{V}(m_\pi^2)\,,
\nonumber\\
F_\pi^4 \Delta^V\! m^{2(6)}_\pi&=&
       m_\pi^4{A}^{V}(m_\pi^2)\, \Big(  - l_4^r + 5\,l_3^r
      + 8\,l_2^r + 14\,l_1^r \Big)
       + m_\pi^2 {A}_{23}^{V}(m_\pi^2) \,
          \Big(  - 12\,l_2^r - 6\,l_1^r \Big)
\nonumber\\&&
       + {A}^{V}(m_\pi^2)\, \Big( 13/12\,\frac{1}{16\pi^2}\,m_\pi^4 - 7/4\,\overline{A}(m_\pi^2)\,m_\pi^2 \Big)
       + {A}^{V}(m_\pi^2)^2\, \Big(  - 3/8\,m_\pi^2 \Big)
\nonumber\\&&
       + {A}^{V}(m_\pi^2)\,{B}^{0V}(m_\pi^2) \, \Big( 1/4\,m_\pi^4 \Big)
       + {H}^V(m_\pi^2,m_\pi^2,m_\pi^2,m_\pi^2) \, \Big( 5/6\,m_\pi^4 \Big)
\nonumber\\&&
       + {H}_{21}^V(m_\pi^2,m_\pi^2,m_\pi^2,m_\pi^2) \, \Big( 3\,m_\pi^4 \Big)
       + {H}_{27}^V(m_\pi^2,m_\pi^2,m_\pi^2,m_\pi^2) \, \Big(  - 3\,m_\pi^2 \Big)\,.
\ea
$\Delta^V\! m^{2(4)}_\pi$ agrees with the results of 
\cite{Gasser:1986vb}.
The comparison of $\Delta^V m^{2(6)}_\pi$ with the result in
\cite{Colangelo:2006mp} is not quite so simple. The reason is that the
splitting in parts has been done very differently there and here.
However, we agree on the sunset part, (44) in \cite{Colangelo:2006mp}
and on the part that has $l_i^r$ multiplying finite volume integrals
in (38) in \cite{Colangelo:2006mp}. The latter was first derived
in \cite{Colangelo:2003hf}.
Both their and our result are independent of the subtraction scale.

The pion decay constant is defined by
\be
\label{defFpinf2}
\langle0|\bar u\gamma_\mu\gamma_5 d|\pi^-(p)\rangle = \sqrt{2}iF_\pi p_\mu\,.
\ee
It can be computed by the diagrams of Fig.~\ref{figdiagrams} where the outgoing
meson is replaced by an insertion of the axial current. The diagrams needed for
wave-function renormalization are the same as those for the mass.
The calculation proceeds along the same lines as above. We reproduce the
known infinite volume results of
\cite{Burgi:1996qi,Bijnens:1995yn,Bijnens:1998fm}. The decay constant at
finite volume we write as
\be
F_\pi^V = F_\pi+\Delta^V\!F_\pi\,,
\quad
\Delta^V\!F_\pi =  \Delta^V\!F_\pi^{(4)}+\Delta^V\!F_\pi^{(6)}\,.
\ee
The results are:
\ba
F_\pi \Delta^V\!F_\pi^{(4)} &=& 
        {A}^{V}(m_\pi^2)\,,
\nonumber\\
F_\pi^3 \Delta^V\!F_\pi^{(6)} &=& 
       + {A}^{V}(m_\pi^2)m_\pi^2 \, \Big( 3/2\,l_4^r - 4\,l_2^r - 7\,l_1^r \Big)
       + {A}_{23}^{V}(m_\pi^2) \, \Big( 6\,l_2^r + 3\,l_1^r \Big)
\nonumber\\&&
       + {A}^{V}(m_\pi^2) \, \Big(  - 1/3\,\frac{1}{16\pi^2}\,m_\pi^2 + 1/2\,\overline{A}(m_\pi^2) \Big)
       + {A}^{V}(m_\pi^2)\,{B}^{0V}(m_\pi^2) \, \Big(  - 1/2\,m_\pi^2 \Big)
\nonumber\\&&
       + {H}^V(m_\pi^2,m_\pi^2,m_\pi^2,m_\pi^2) \, \Big(  - 1/2\,m_\pi^2 \Big)
       + {H}_{27}^V(m_\pi^2,m_\pi^2,m_\pi^2,m_\pi^2) \, \Big( 3/2 \Big)
\nonumber\\&&
       + {H}^{V\prime}(m_\pi^2,m_\pi^2,m_\pi^2,m_\pi^2) \, \Big( 5/12\,m_\pi^4 \Big)
       + {H}_{21}^{V\prime}(m_\pi^2,m_\pi^2,m_\pi^2,m_\pi^2) \, \Big( 3/2\,m_\pi^4 \Big)
\nonumber\\&&
       + {H}_{27}^{V\prime}(m_\pi^2,m_\pi^2,m_\pi^2,m_\pi^2) \, \Big(  - 3/2\,m_\pi^2 \Big)\,.
\ea 
$\Delta^V\!F_\pi^{(4)}$ agrees with the results of 
\cite{Gasser:1986vb}. Here there exists no full
two-loop calculation but an evaluation for the case with at most one propagator
at finite volume \cite{Colangelo:2004xr}. We agree with their result for the
terms containing $l_i^r$ if the term multiplying $B^2$ in (54) in that paper is
divided by 2. Comparing with the remainder is difficult due to the very
different treatment of the loop integrals.

\section{Three-flavour results}
\label{sec:nf3}

The principle of the calculation is exactly the same as before.
The diagrams needed for the mass are shown in Fig.~\ref{figdiagrams}.
However, we now need to use the three-flavour Lagrangians and include the
kaons and eta as well. As a result the expressions become much more cumbersome.
Here we use as symbols, $m_\pi$, $m_K$ and $m_\eta$ as the physical
volume pion, kaon and eta mass at infinite volume.
We have rewritten all expressions as an expansion in these masses and
in the physical pion decay constant at infinite volume.
Given that the eta mass to lowest order is given by the Gell-Mann--Okubo
relation, there is an inherent ambiguity in precisely how one writes the
result in the combination of kaon and eta masses. The form of the $p^6$ result
given here is to be used together with the form for the $p^4$ expressions
given here as well. 

The pion, kaon and eta masses at two-loop order in infinite volume
are known, \cite{Amoros:1999dp}, we have reproduced that result.
The finite volume corrections for the masses are given by
\be
m_i^{V2} = m_i^2+\Delta^V\! m^2_i\,,
\quad \Delta^V\! m^2_i = \Delta^V\! m^{2(4)}_i+\Delta^V\! m^{2(6)}_i\,,
\ee
for $i=\pi,K,\eta$.
The $p^4$ results are:
\ba
\label{minf3p4}
F_\pi^2 \Delta^V\! m^{2(4)}_\pi
&=& 
        {A}^{V}(m_\pi^2) \, \Big(  - 1/2\,m_\pi^2 \Big)
       + {A}^{V}(m_\eta^2) \, \Big( 1/6\,m_\pi^2 \Big)\,,
\nonumber\\
F_\pi^2 \Delta^V\! m^{2(4)}_K &=&
       {A}^{V}(m_\eta^2) \, \Big(  - 1/4\,m_\eta^2 - 1/12\,m_\pi^2 \Big)\,,
\nonumber\\
F_\pi^2 \Delta^V\! m^{2(4)}_\eta &=&
        {A}^{V}(m_\pi^2) \, \Big( 1/2\,m_\pi^2 \Big)
       + {A}^{V}(m_K^2) \, \Big(  - m_\eta^2 - 1/3\,m_\pi^2 \Big)
\nonumber\\&&
       + {A}^{V}(m_\eta^2) \, \Big( 8/9\,m_K^2 - 7/18\,m_\pi^2 \Big)\,.
\ea
These agree with the expressions in \cite{Becirevic:2003wk,DescotesGenon:2004iu,Colangelo:2005gd}. The way in
which the corrections are written is to be in agreement with the way the
infinite volume result was written in \cite{Amoros:1999dp}.
The order $p^6$ expressions are rather large, they can be
found in App.~\ref{appmass}. The contributions with at most one pion propagator
at finite volume were calculated in \cite{Colangelo:2005gd} for the kaon and
eta in three flavour ChPT, the expression for the pion was done in
two-flavour ChPT and discussed above. We agree with the $L_i^r$ times finite
volume part there. The remainder is difficult to compare due to the different
treatment of the integrals.

The decay constants for the mesons are defined similarly to (\ref{defFpinf2})
via
\ba
\label{defFpi}
\langle0|\bar u\gamma_\mu\gamma_5 d|\pi^-(p)\rangle &=& \sqrt{2}iF_\pi p_\mu\,,
\nonumber\\
\langle0|\bar u\gamma_\mu\gamma_5 s|K^-(p)\rangle &=& \sqrt{2}iF_K p_\mu\,,
\nonumber\\
\langle0|\frac{1}{\sqrt{6}}\left(\bar u\gamma_\mu\gamma_5 u+\bar d\gamma_\mu\gamma_5 d-2\bar s\gamma_\mu\gamma_5 s\right)|\eta(p)\rangle &=& \sqrt{2}iF_\eta p_\mu\,.
\ea
Note that since we work in the isospin limit, we use the octet axial current to
define the eta decay constant.

We define
\be
F_i^{V} = F_i+\Delta^V\! F_i\,,
\quad \Delta^V\! F_i = \Delta^V\! F^{(4)}_i+\Delta^V\! F^{(6)}_i\,,
\ee
for $i=\pi,K,\eta$.
The pion, kaon and eta decay constants at two-loop order in infinite volume
are known, \cite{Amoros:1999dp}, we have reproduced that result.
Note that we give the corrections to the decay constants here, not divided by
the chiral limit decay constant as in \cite{Amoros:1999dp}.
Note the correction for the expressions for the infinite volume decay constants
described in the erratum of \cite{Amoros:2000mc}. The correct expressions
can be downloaded from \cite{homepage}

The order $p^4$ results are
\ba
\label{Finf3p4}
F_\pi\Delta^V\!F_\pi^{(4)}&=&
        {A}^{V}(m_\pi^2)
       + {A}^{V}(m_K^2) \, \Big( 1/2 \Big)\,,
\nonumber\\
F_\pi\Delta^V\!F_K^{(4)}&=&
        {A}^{V}(m_\pi^2) \, \Big( 3/8 \Big)
       + {A}^{V}(m_K^2) \, \Big( 3/4 \Big)
       + {A}^{V}(m_\eta^2) \, \Big( 3/8 \Big)\,,
\nonumber\\
F_\pi\Delta^V\!F_\eta^{(4)}&=&
        {A}^{V}(m_K^2) \, \Big( 3/2 \Big)\,.
\ea
These agree with  \cite{Becirevic:2003wk,DescotesGenon:2004iu,Colangelo:2005gd}.
The $p^6$ expressions are again rather long and are
given in App.~\ref{appdecay}.
The contributions with at most one-pion propagator
at finite volume were calculated in \cite{Colangelo:2005gd} for the kaon
in three flavour ChPT, the expression for the pion was done in
two-flavour ChPT and discussed above. 
We agree with the $L_i^r$ dependent part if we multiply the contribution
from the term with $B^2$ in (57) in by $1/2$. This is the same factor we needed
to get agreement for the two-flavour pion decay constant.

\section{Numerical results}
\label{numerics}

For numerical input we use $F_\pi= 92.2$~MeV,
$m_\pi=m_{\pi^0}= 134.9764$~MeV, the average $m_K$ with electromagnetic
 effects removed with the estimate of \cite{Bijnens:1996kk},
$m_K = 494.53$~MeV, and $m_\eta= 547.30$~MeV.
The values of the low-energy constants, we take from the last review
\cite{Bijnens:2014lea}. We always use a subtraction scale $\mu = 770$~MeV.

\subsection{Two-flavour results}

\begin{figure}[tb!]
\begin{minipage}{0.49\textwidth}
\includegraphics[width=0.99\textwidth]{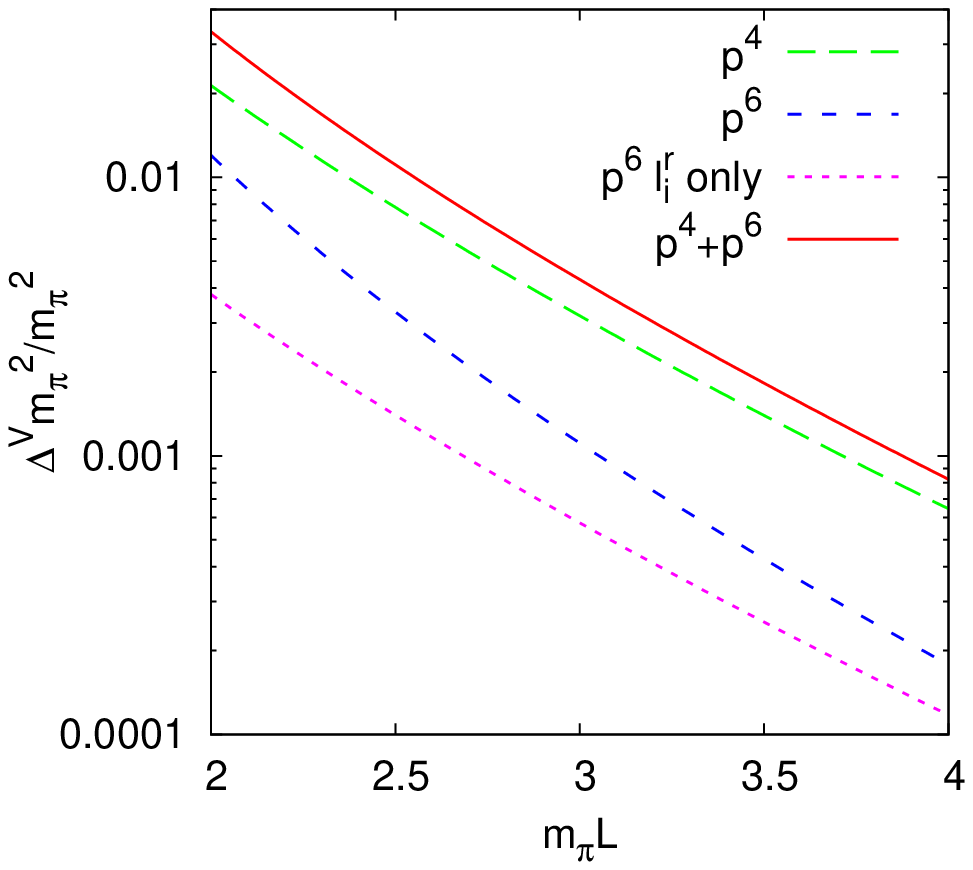}
\centerline{(a)}
\end{minipage}
\begin{minipage}{0.49\textwidth}
\includegraphics[width=0.99\textwidth]{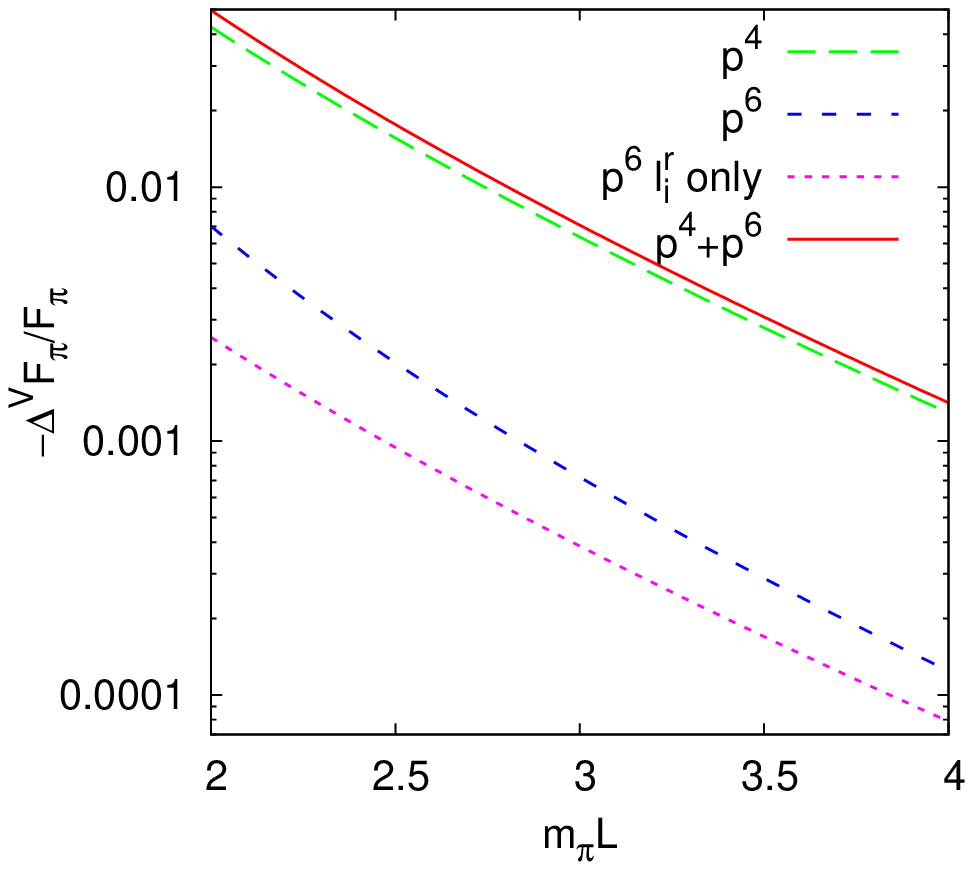}
\centerline{(b)}
\end{minipage}
\caption{The relative finite volume corrections for the mass squared and decay
constant of the pion in two-flavour ChPT at a fixed infinite volume pion mass
$m_\pi=m_{\pi^0}$. Shown are the one-loop or $p^4$ corrections,
the full $p^6$ result and the part only dependent on the $l_i^r$, $p^6 l_i^r$,
and the sum of the $p^4$ and $p^6$ result.
$m_\pi L=2,4$ correspond to $L\approx 2.9,5.8$~fm.
(a) The pion mass, plotted is $(m_\pi^{V2}-m_\pi^2)/m_\pi^2$.
(b) The pion decay constant. Plotted is $-(F_\pi^V-F_\pi)/F_\pi$.}
\label{figmassdecay}
\end{figure}

The $l_i^r$ we use we define via the usual $\bar l_i$ defined at the scale
of the charged pion mass. The actual values we use are
$\bar l_1=-0.4, \bar l_2=4.3,\bar l_3=3.0,\bar l_4=4.3$.
The relative finite volume corrections to $m_\pi^2$ are shown in
Fig.~\ref{figmassdecay}(a) as a function of $m_\pi L$. We have checked
that changing the scale to $\mu=500$~MeV does not change the result, but it
does increase the $l_i^r$ part.
The equivalent plot for the relative correction to $F_\pi$ is shown in Fig.~\ref{figmassdecay}(b).

\begin{figure}[tb!]
\begin{minipage}{0.49\textwidth}
\includegraphics[width=0.99\textwidth]{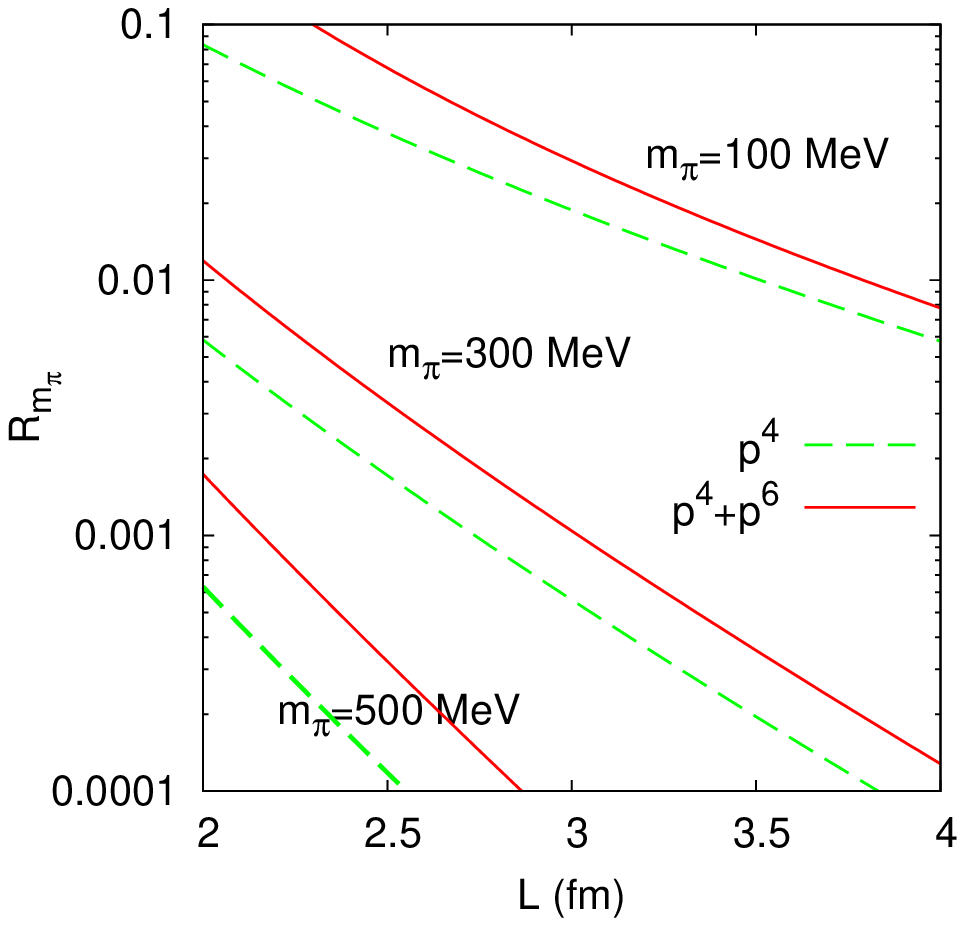}
\centerline{(a)}
\end{minipage}
\begin{minipage}{0.49\textwidth}
\includegraphics[width=0.99\textwidth]{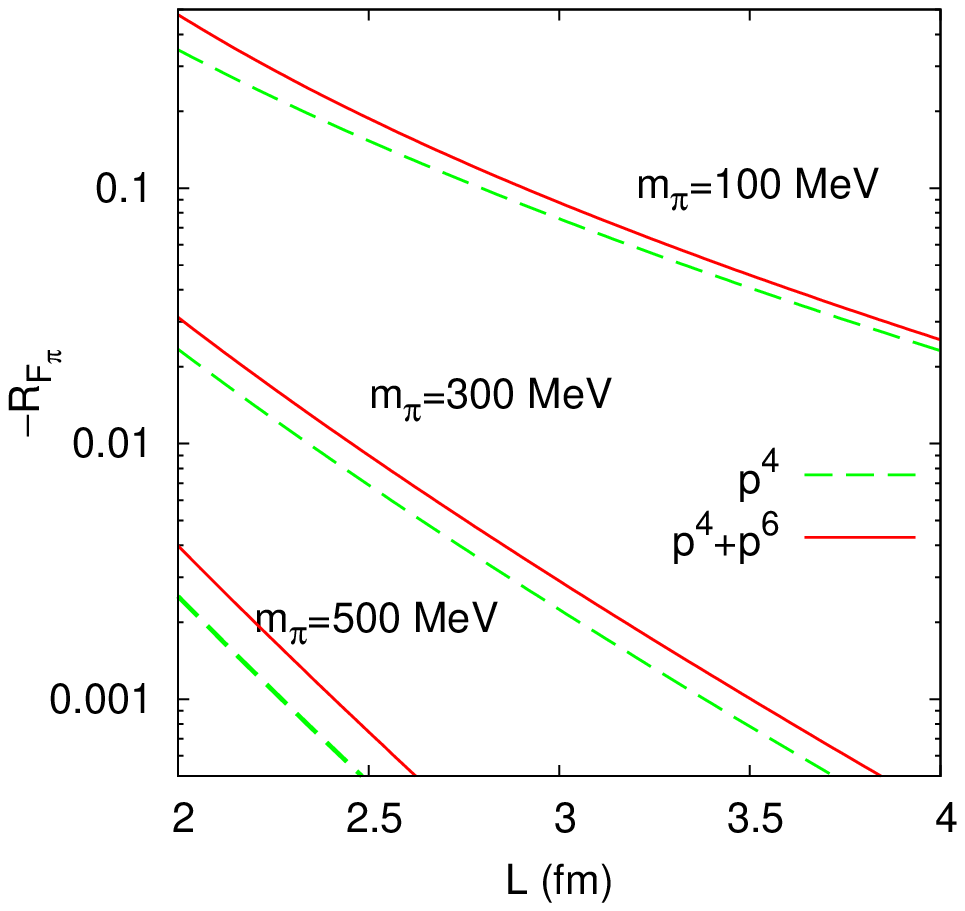}
\centerline{(b)}
\end{minipage}
\caption{The relative finite volume corrections for the mass and decay
constant of the pion in two-flavour ChPT at three values of the infinite
volume pion mass.
(a) $R_{m_\pi}=m_\pi^V/m_\pi-1$. (b) $R_{F_\pi}=F_\pi^V/F_\pi-1$, plotted is $-R_{F_\pi}$.}
\label{figrmpiL}
\end{figure}

We can also perform a study of the corrections at other values of $m_\pi$ or
as a function of $m_\pi$. One of the problems here is what to with the
value of $F_\pi$ that should be used. If we use the infinite volume formulas
to two-loop order of \cite{Bijnens:1998fm} which are expressed in the
form $F_\pi/F = f(F_\pi,m_\pi)$ for another pion mass $\tilde m_\pi$ we
determine the associated value of the decay constant, $\tilde F_\pi$ by
solving $\tilde F_\pi/F_\pi = f(\tilde F_\pi,\tilde m_\pi)/f(F_\pi,m_\pi)$
numerically. The contribution from the $p^6$ LECs $c_i^r$ we have put to zero.
This procedure might differ from the values of $\tilde F_\pi$ used
in \cite{Colangelo:2006mp}. To compare with their numerical results
we have plotted in Fig.~\ref{figrmpiL} the equivalent of their Fig.~5.
Namely $R_{m_\pi}=m_\pi^V/m_\pi-1$ where we have numerically calculated
$R_{m_\pi}=\sqrt{(m_\pi^2+\Delta^V\! m_\pi^2)/m_\pi^2}-1$. 
The calculated values of $F_\pi$ are $90.1,103.2,113.8$ for
$m_\pi=100,300,500$~MeV. The resulting values of $R_{m_\pi}$ as shown in
Fig.~\ref{figrmpiL}(a) are in reasonable agreement with Fig.~5
in \cite{Colangelo:2006mp}. There is already a difference at order $p^4$, so
we suspect it is simply due to somewhat different values of $F_\pi$.
The one-loop result for $R_{F_\pi}$ agrees with Fig.~2 in
\cite{Colangelo:2004xr} with small differences probably due to the
difference in $F_\pi$ and the difference in the $l_i^r$-dependent part.
Our result for the $p^6$ result is somewhat larger.

\subsection{Three-flavour results: masses}

The values of the low-energy constants, $L_i^r$ and $C_i^r$, we take from the
review \cite{Bijnens:2014lea}, in particular the set labeled BE14 there.
In addition, the formulas require the infinite volume physical masses
for the pion, kaon and eta mass as well as the pion decay constant.
The masses and $F_\pi$ we use for the physical isospin averaged case are listed
at the start of this section. For changed values of the infinite volume
pion and kaon mass, $\tilde m_\pi,\tilde m_K$, we proceed similarly to
$F_\pi$ for the two-flavour case. We solve self-consistently the set of
equations for $\tilde m_\eta$, $\tilde F_\pi$, $\tilde F_K/\tilde F_\pi$
and $\tilde F_\eta/\tilde F_\pi$. For the latter ratios we use the expanded
version, similar to what was done in \cite{Bijnens:2014lea}, see Eq.~(45)
in there.
The results for a number of
input cases is shown in Tab.~\ref{tabmpimk}. 
\begin{table}[tb!]
\centerline{\begin{tabular}{|ccccccccc|}
\hline
$m_\pi$ & $m_K$ & $m_\eta$ & $F_\pi$ & $F_K/F_\pi$ & $F_\eta/F_\pi$ &
 $\hat m/\hat m_{\mathrm{phys}}$ & $m_s/m_{s\mathrm{phys}} $ & $m_s/\hat m$\\
\hline
134.9764$^*$ & 494.53$^*$ & 545.9 & 92.2$^*$ & 1.199 & 1.306 & 1$^*$ & 1$^*$ & 27.3\\
100 & 487.14 & 540.46 & 90.4 & 1.219 & 1.337 & 0.547 & 1.000 & 49.9\\
300 & 549.6  & 593.73 & 101.4& 1.099 & 1.154 & 5.025 & 1.000 & 5.43\\
100 & 400    & 446.53 & 87.3 & 1.199 & 1.293 & 0.518 & 0.644 & 33.9\\
100 & 495    & 549.07 & 90.7 & 1.219 & 1.340 & 0.550 & 1.037 & 51.4\\
300 & 495    & 533.00 & 100.3& 1.094 & 1.138 & 4.867 & 0.778 & 4.36\\
495 & 495    & 495.00 & 108.0& 1     & 1     &12.70  & 0.465 & 1\\
\hline
\end{tabular}}
\caption{The self consistent solution for the infinite volume values of
$m_\eta$, $F_\pi$, $F_K$, $F_\eta$ and the output quark mass ratios compared with the physical one. Units for dimensional quantities are in $MeV$.
The input values for the physical case are starred.}
\label{tabmpimk}
\end{table}
The top line is the physical case
The resulting output is within the expected quality of the fit in
\cite{Bijnens:2014lea}. The next two lines have the kaon mass tuned to keep
the same value of $m_s$. The value of $F_\pi$ can be compared with
the result for the two-flavour case given above.

\begin{figure}[tb!]
\begin{minipage}{0.49\textwidth}
\includegraphics[width=0.99\textwidth]{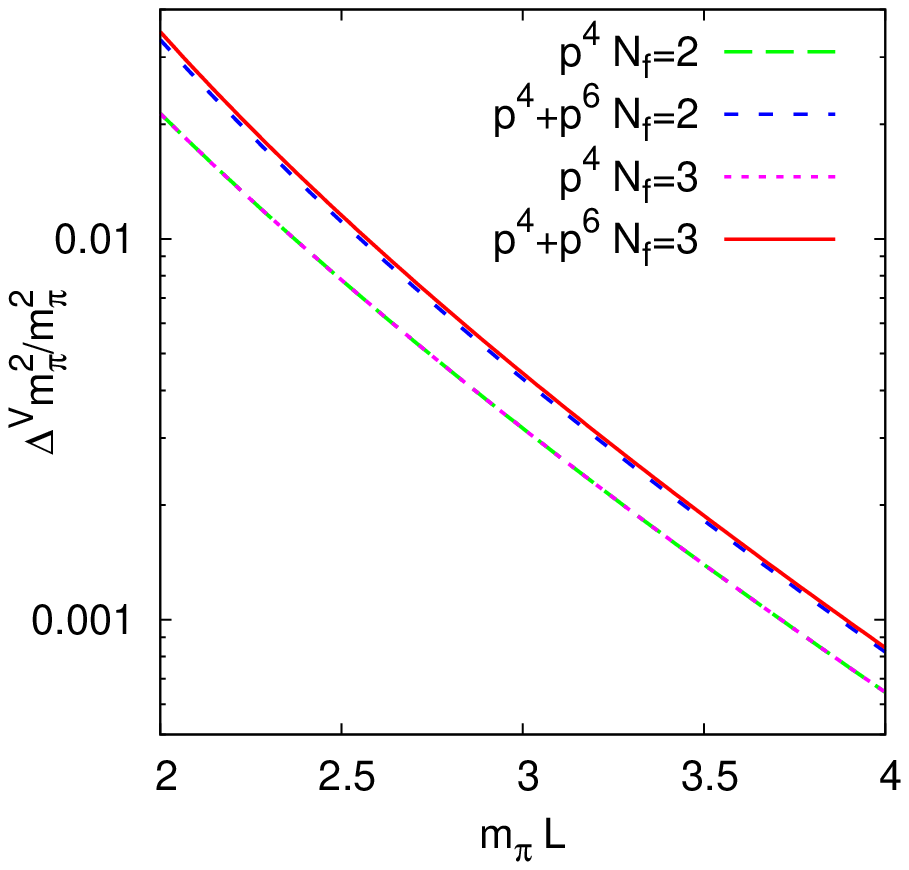}
\centerline{(a)}
\end{minipage}
\begin{minipage}{0.49\textwidth}
\includegraphics[width=0.99\textwidth]{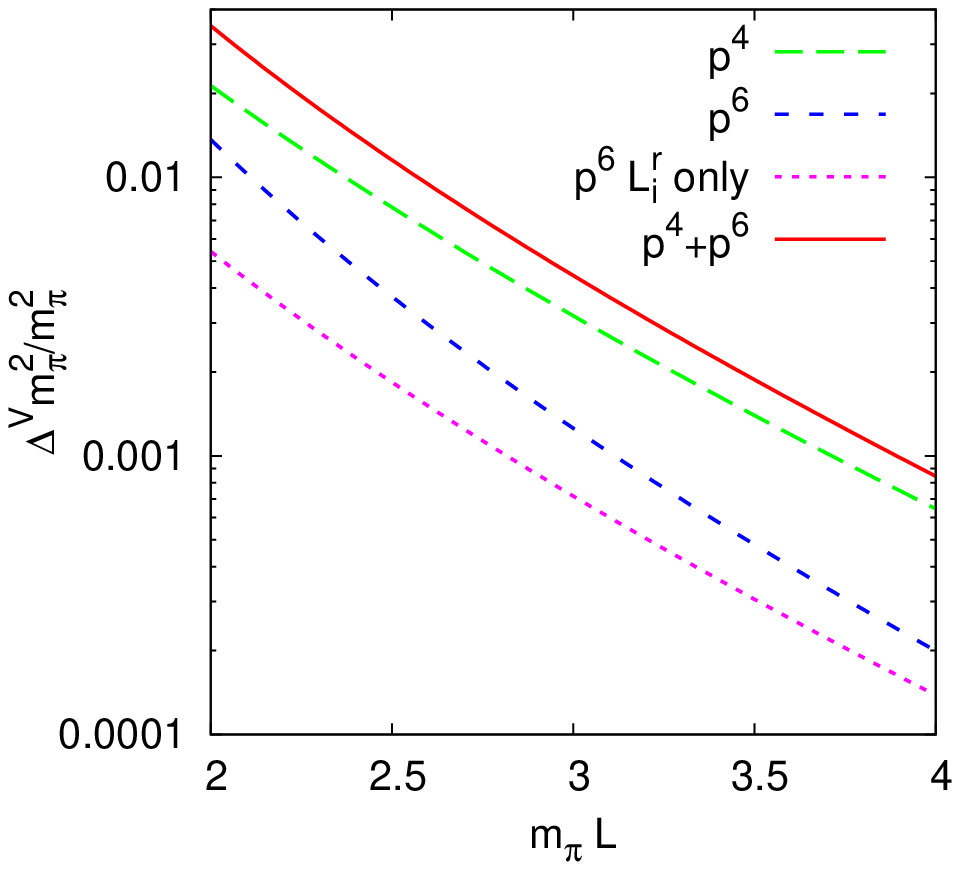}
\centerline{(b)}
\end{minipage}
\caption{The finite volume corrections to the pion mass squared
at $m_\pi=m_{\pi^0}$. All other inputs are given in the text. 
Plotted is the quantity $(m_\pi^{V2}-m_\pi^2)/m_\pi^2$.
 (a) Comparison of the two- and three-flavour ChPT results. 
(b) The corrections for the three-flavour case also showing
the $L_i^r$ dependent part.}
\label{figmpi}
\end{figure}

Let us have a look at the pion mass finite volume corrections for the physical
case. 
The comparison of the
two- and three-flavour results are plotted in Fig.~\ref{figmpi}(a).
The one-loop result differs only by a very small kaon and eta loop. The
difference is not visible in the figure. The two-loop results are also
in very good agreement. The convergence is quite reasonable.

\begin{figure}[tb!]
\begin{minipage}{0.49\textwidth}
\includegraphics[width=0.99\textwidth]{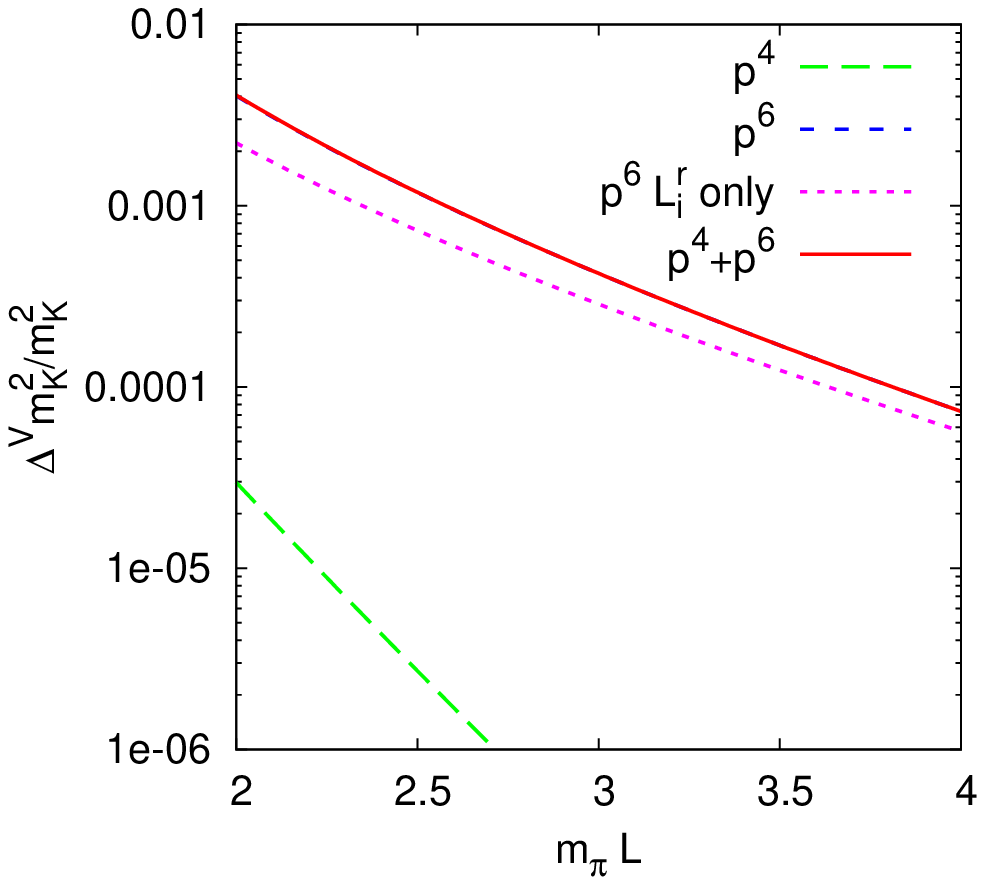}
\centerline{(a)}
\end{minipage}
\begin{minipage}{0.49\textwidth}
\includegraphics[width=0.99\textwidth]{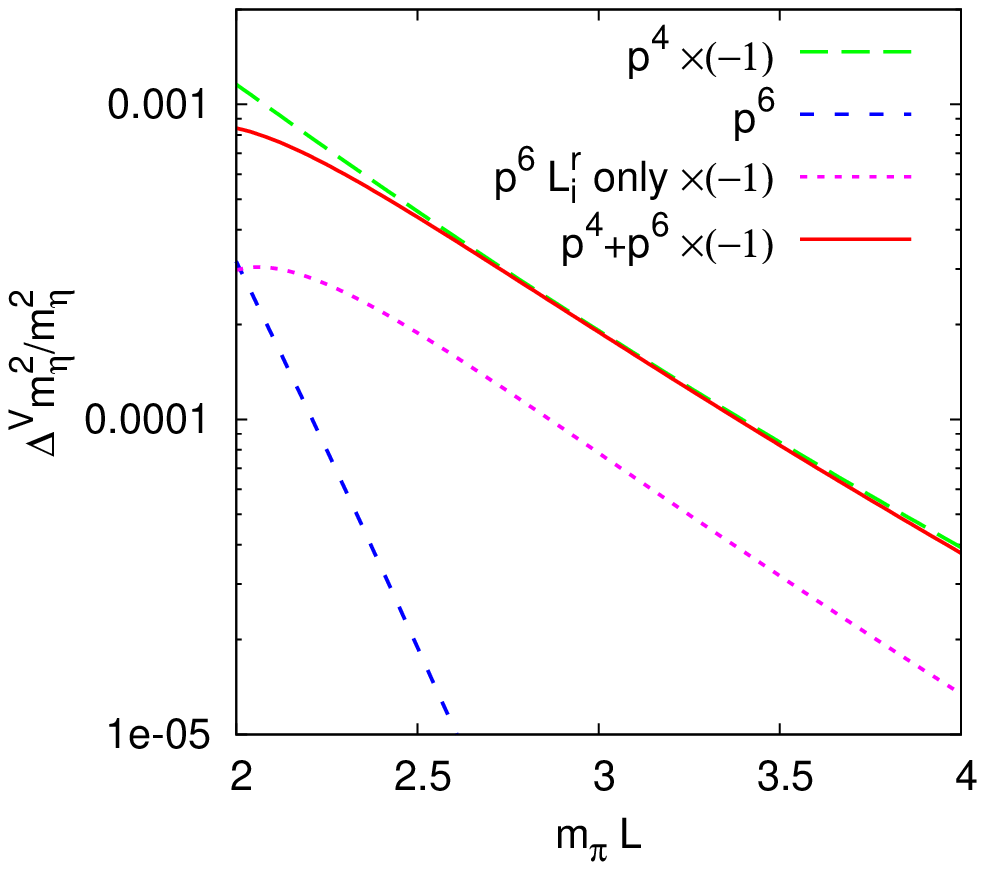}
\centerline{(b)}
\end{minipage}
\caption{The corrections to the kaon and eta mass squared for the physical case.
Plotted is the quantity $(m_i^{V2}-m_i^2)/m_i^2$ for $i=K,\eta$.
Shown are the one-loop, the two-loop, the sum and the two-loop $L_i^r$
dependent part.
 (a) Kaon, the $p^4$ is so small that $p^6$ and $p^4+p^6$ are indistinguishable. (b) Eta, note the signs, some parts are negative.}
\label{figmketa}
\end{figure}

The equivalent results for the kaon and eta are plotted in Fig.~\ref{figmketa}.
The one-loop result for the kaon mass has only an eta loop
as can be seen from (\ref{minf3p4}). As a result,
that part is very small. The total result is thus essentially coming only from
two-loop order. The eta mass has a negative one-loop finite volume
contribution. The pure loop part and the $L_i^r$-dependent part of the $p^6$
contribution are of the expected size. However, there is a very strong
cancellation between the two parts leaving a very small positive correction.
The total finite volume correction for the eta mass in negative.

\begin{figure}[tb!]
\begin{minipage}{0.49\textwidth}
\includegraphics[width=0.99\textwidth]{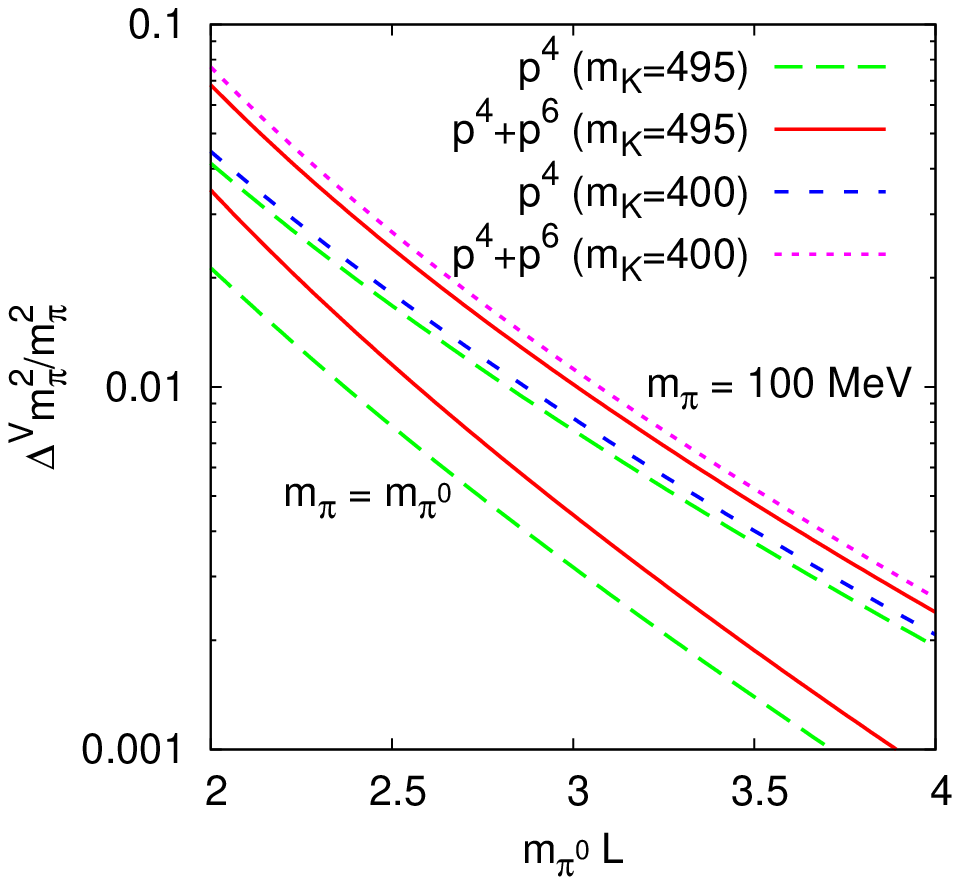}
\centerline{(a)}
\end{minipage}
\begin{minipage}{0.49\textwidth}
\includegraphics[width=0.99\textwidth]{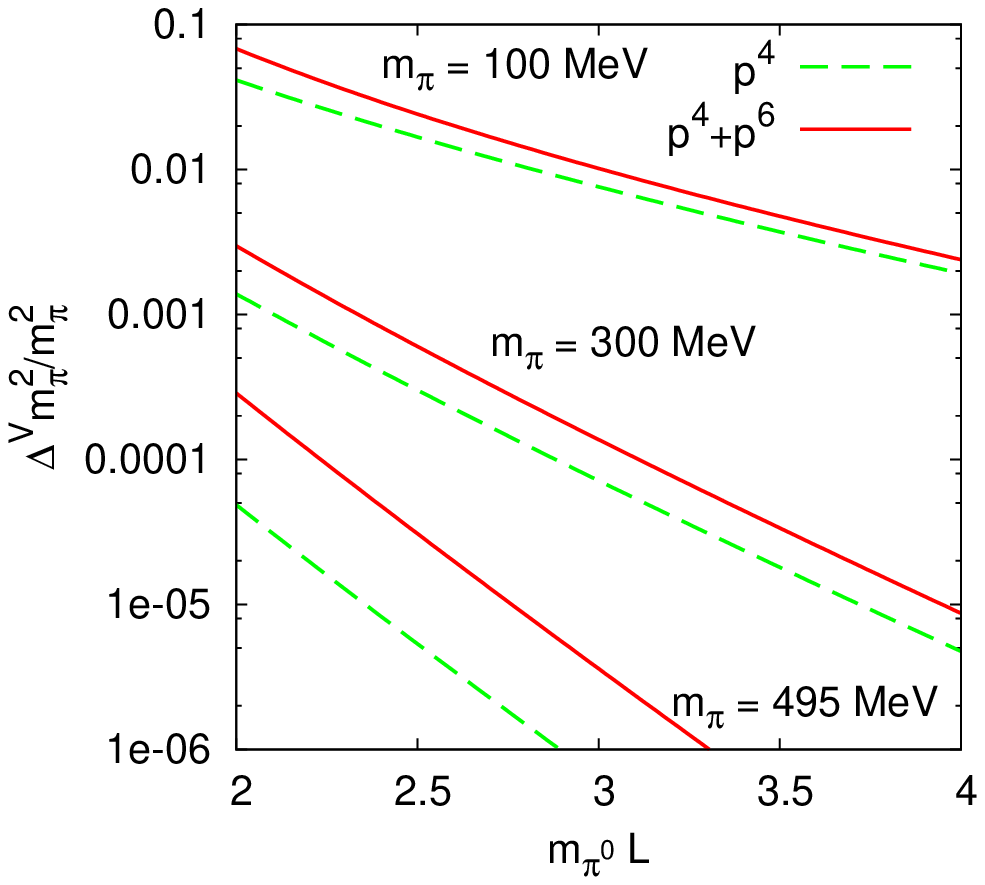}
\centerline{(b)}
\end{minipage}
\caption{The finite volume corrections to the pion mass squared
for a number of cases listed in Tab.~\ref{tabmpimk}.
Plotted is the quantity $(m_\pi^{V2}-m_\pi^2)/m_\pi^2$.
(a) Physical case, bottom two lines, $(m_\pi,m_K) = (100,495)$ and
 $(100,400)$~MeV.
(b) $m_K=495$~MeV and $m_\pi=100,300,495$~MeV.
The size $L$ is given in units of the physical $\pi^0$ mass.}
\label{figmpi2}
\end{figure}

We can also check how the finite volume correction depends on the different
masses. In Fig.~\ref{figmpi2} we have plotted the corrections to the
pion mass squared for a number of different scenarios. In Fig.~\ref{figmpi2}(a)
we look at three cases. The bottom two lines are the physical case
labeled with $m_\pi=m_{\pi^0}$ while the top four lines are with
$m_\pi=100$~MeV. There we have plotted two cases, $m_K=400$ and $495$~MeV.
The effect of the change in the pion mass is quite large while the
effect due to the kaon mass change is smaller.
The effect of changing the pion mass can be better seen
in Fig.~\ref{figmpi2}(b) where we kept the kaon mass at 495~MeV while
varying the pion mass.
The $L$ dependence is given as a function of $m_{\pi^0}L$ with the
physical $\pi^0$ mass.

\begin{figure}[tb!]
\begin{minipage}{0.49\textwidth}
\includegraphics[width=0.99\textwidth]{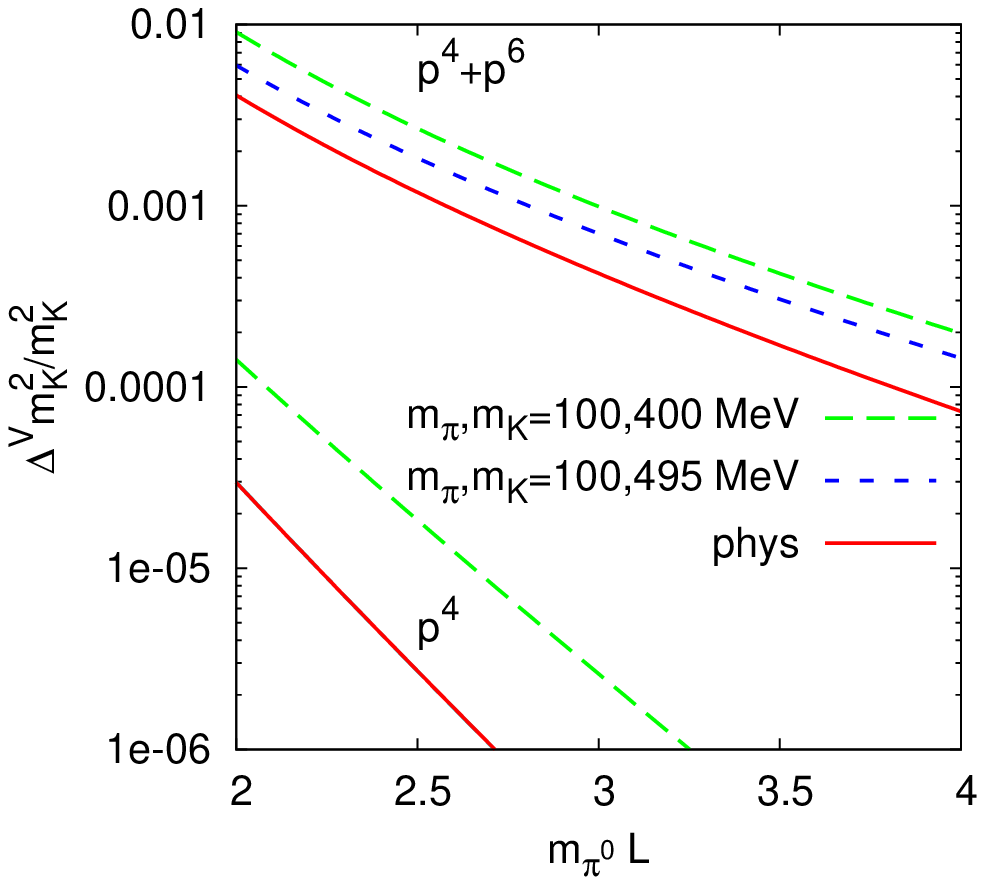}
\centerline{(a)}
\end{minipage}
\begin{minipage}{0.49\textwidth}
\includegraphics[width=0.99\textwidth]{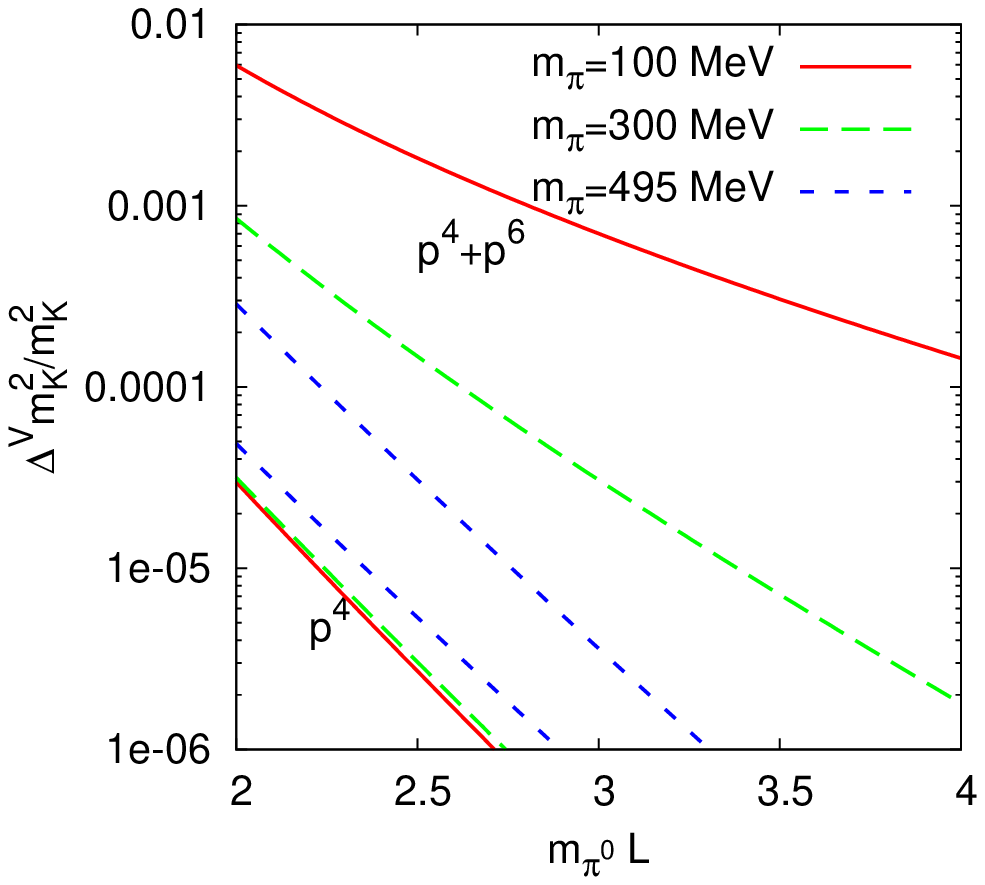}
\centerline{(b)}
\end{minipage}
\caption{The finite volume corrections to the kaon mass squared
for a number of cases listed in Tab.~\ref{tabmpimk}.for the physical case.
Plotted is the quantity $(m_K^{V2}-m_K^2)/m_K^2$.
(a) Physical case and $(m_\pi,m_K) = (100,495)$ and $(100,400)$~MeV.
(b) $m_K=495$~MeV and $m_\pi=100,300,495$~MeV.
The size $L$ is given in units of the physical $\pi^0$ mass.}
\label{figmk2}
\end{figure}

We have plotted the same cases for the finite volume corrections to the kaon
mass squared in Fig.~\ref{figmk2}.
The one-loop correction for the physical case and $m_\pi,m_K=100,495$~MeV
is virtually identical. The $p^4+p^6$ is a bit more different for the three
cases as can be seen in Fig.~\ref{figmk2}(a). In Fig.~\ref{figmk2}(b) we have
shown the corrections for a fixed kaon mass but three different pion masses.
The bottom three lines are the one-loop result while the top three
lines are the full result.
Note that, as it should be, the case where the pion mass and kaon mass
are the same,
the finite volume corrections to the kaon are the same as for the pion
in Fig.~\ref{figmpi2}(b). This is another small check on our result.

\begin{figure}[tb!]
\begin{minipage}{0.49\textwidth}
\includegraphics[width=0.99\textwidth]{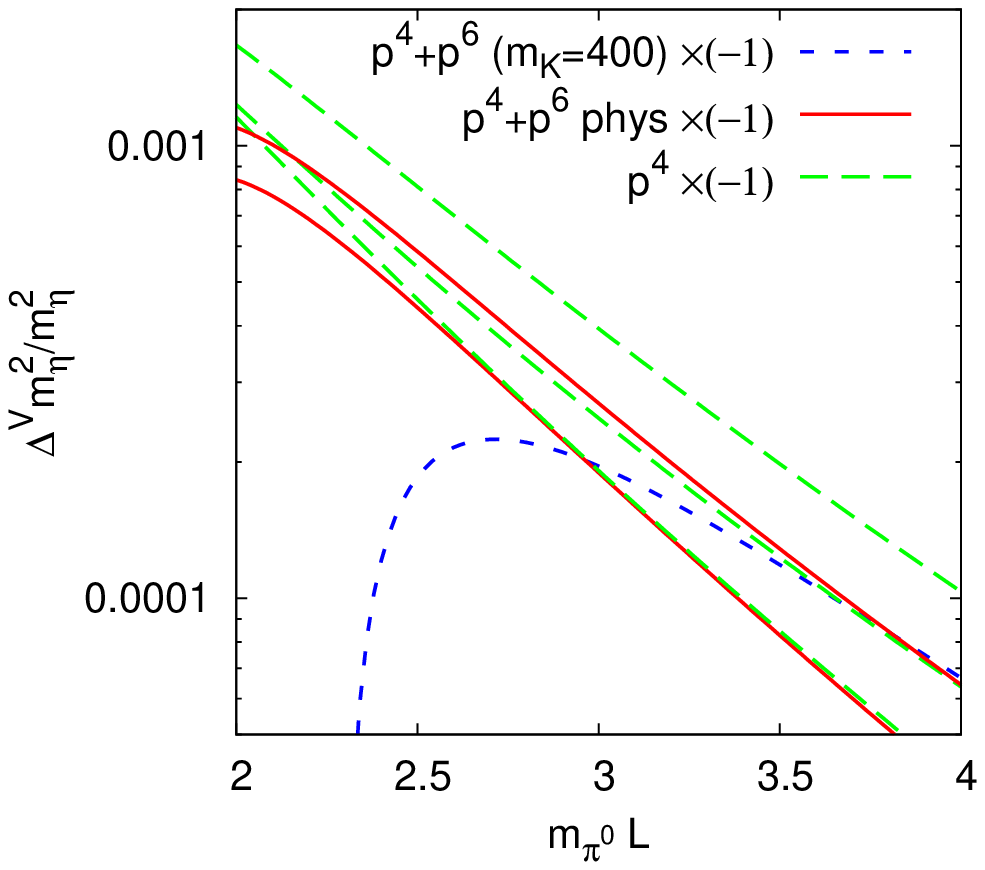}
\centerline{(a)}
\end{minipage}
\begin{minipage}{0.49\textwidth}
\includegraphics[width=0.99\textwidth]{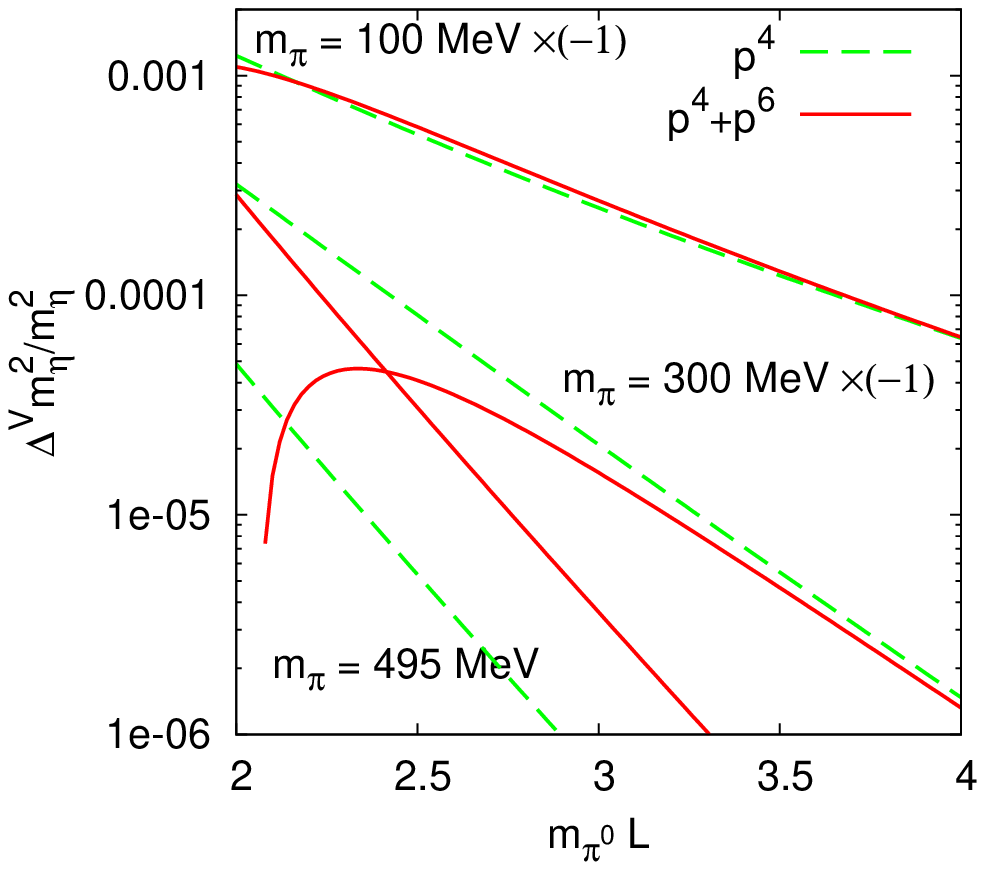}
\centerline{(b)}
\end{minipage}
\caption{The finite volume corrections to the eta mass squared
for a number of cases listed in Tab.~\ref{tabmpimk}.for the physical case.
Plotted is the quantity $(m_\eta^{V2}-m_\eta^2)/m_\eta^2$.
(a) Physical case and $(m_\pi,m_K) = (100,495)$ and $(100,400)$~MeV.
Lines are for the one-loop result at the right bottom physical case, middle $(m_\pi,m_K) = (100,495)$, top  $(m_\pi,m_K) = (100,400)$. The first two have only
a small change due to $p^6$, while for the last case there is a large cancellation between one and two-loops.
(b) $m_K=495$~MeV and $m_\pi=100,300,495$~MeV.
The size $L$ is given in units of the physical $\pi^0$ mass.}
\label{figmeta2}
\end{figure}

We have plotted the same cases once more
for the finite volume corrections to the eta
mass squared in Fig.~\ref{figmeta2}.
Here the result is rather variable due to cancellations.
In Fig.~\ref{figmeta2}(a) the one-loop corrections increase
going from the physical case via $m_\pi,m_K=100,495$~MeV
to $m_\pi,m_K=100,400$~MeV. The two-loop corrections are rather
small in the first two cases, due to the cancellations between the pure two-loop and the $L_i^r$ dependent part.
The one-loop correction for the physical case and $m_\pi,m_K=100,495$~MeV
is virtually identical. The $p^4+p^6$ is a bit more different for the three
cases.
In Fig.~\ref{figmeta2}(b) we have
shown the corrections for a fixed kaon mass but three different pion masses.
The bottom lines are the case with $m_\pi,m_K=495$~MeV. It agrees with the pion and kaon corrections for this case.
For $m_\pi=300$~MeV the correction is negative but goes through zero for small
$L$ due to a cancellation between one-and two-loop results.
The $p^6$ correction for $m_\pi=100$~MeV is very small, we again have a large
cancellation between the pure two-loop and the $L_i^r$ dependent part.

We did not compare with the numerical results in \cite{Colangelo:2005gd},
since there was a small mistake in the relevant figures \cite{colangeloprivate}.

\subsection{Three-flavour results: decay constants}

We will use exactly the same input values as in the previous subsection now but
for the decay constants. Note that here in most cases the finite volume
correction is negative.

\begin{figure}[tb!]
\begin{minipage}{0.49\textwidth}
\includegraphics[width=0.99\textwidth]{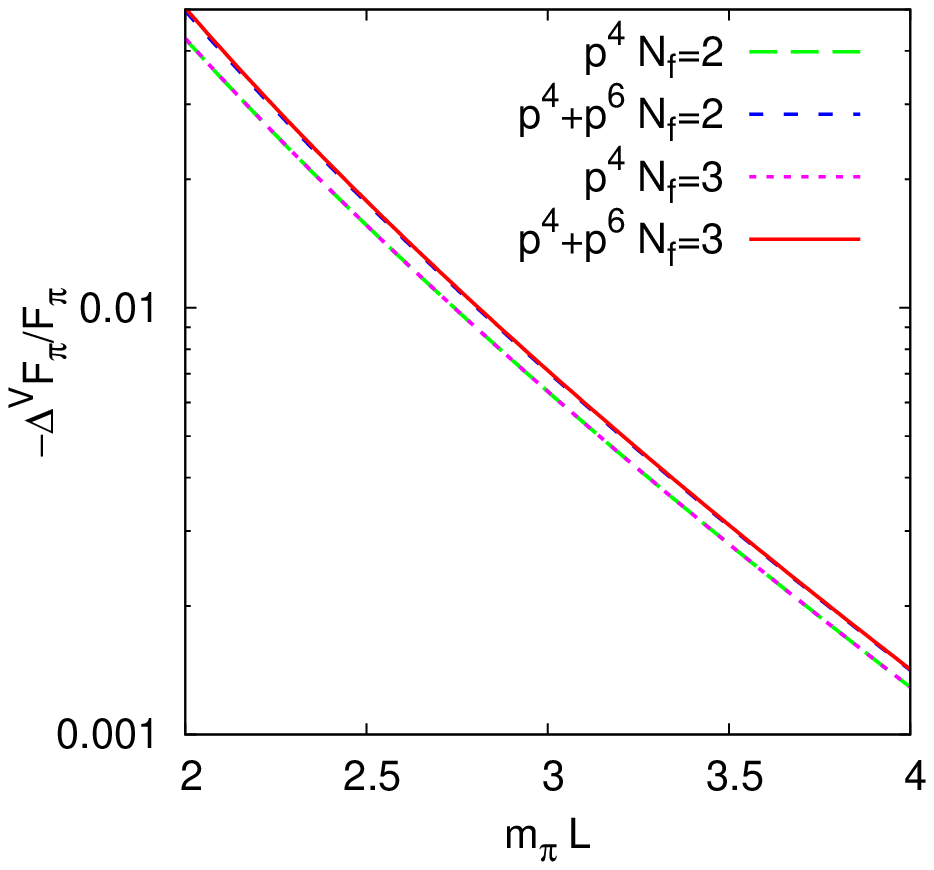}
\centerline{(a)}
\end{minipage}
\begin{minipage}{0.49\textwidth}
\includegraphics[width=0.99\textwidth]{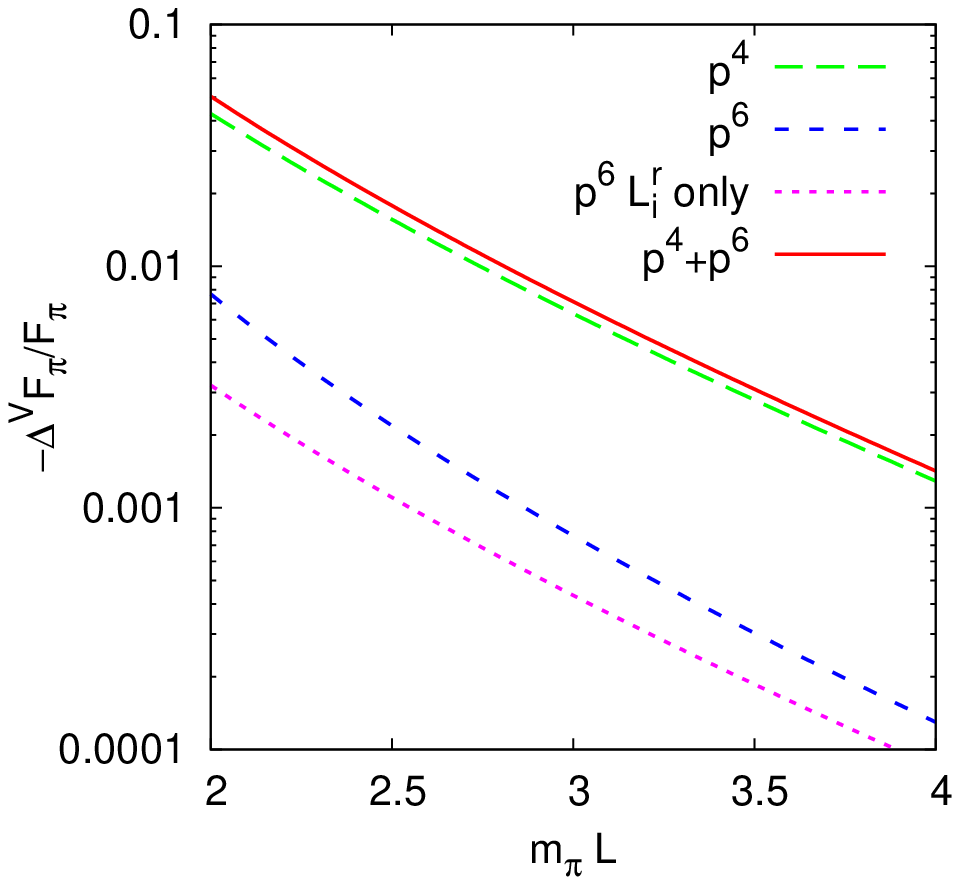}
\centerline{(b)}
\end{minipage}
\caption{The finite volume corrections to the pion decay constant
at $m_\pi=m_{\pi^0}$. All other inputs are given in the text. 
Plotted is the quantity $-(F_\pi^{V}-F_\pi)/F_\pi$.
 (a) Comparison of the two- and three-flavour ChPT results. 
(b) The corrections for the three-flavour case also showing
the $L_i^r$ dependent part.}
\label{figfpi}
\end{figure}

The comparison of the
two- and three-flavour results for the pion
decay constant is plotted in Fig.~\ref{figfpi}(a).
The one-loop result differs only by a very small kaon and eta loop. The
difference is not visible in the figure. The two-loop results are also
essentially indistinguishable. The convergence is quite reasonable.
The bottom line and top line(s) are respectively the one-loop and the sum of
one- and two-loops. Note that in agreement with the earlier estimates there
is a sizable correction at finite volume even at $m_\pi L=2$.

\begin{figure}[tb!]
\begin{minipage}{0.49\textwidth}
\includegraphics[width=0.99\textwidth]{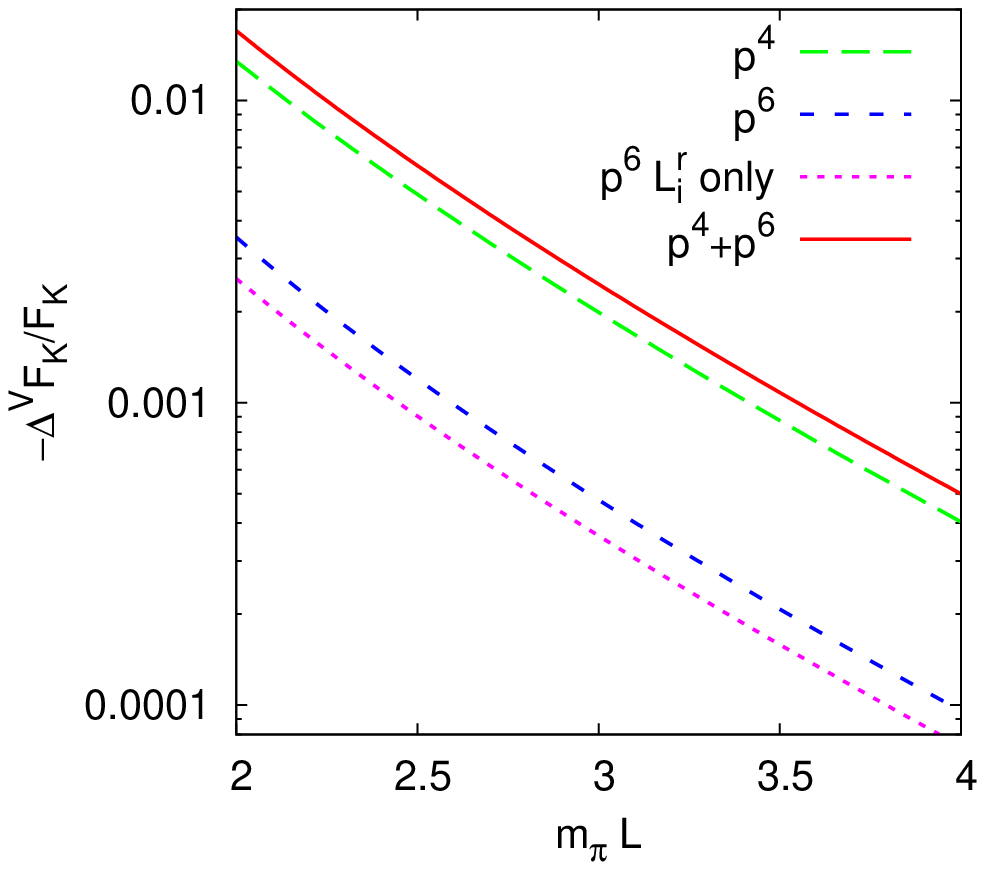}
\centerline{(a)}
\end{minipage}
\begin{minipage}{0.49\textwidth}
\includegraphics[width=0.99\textwidth]{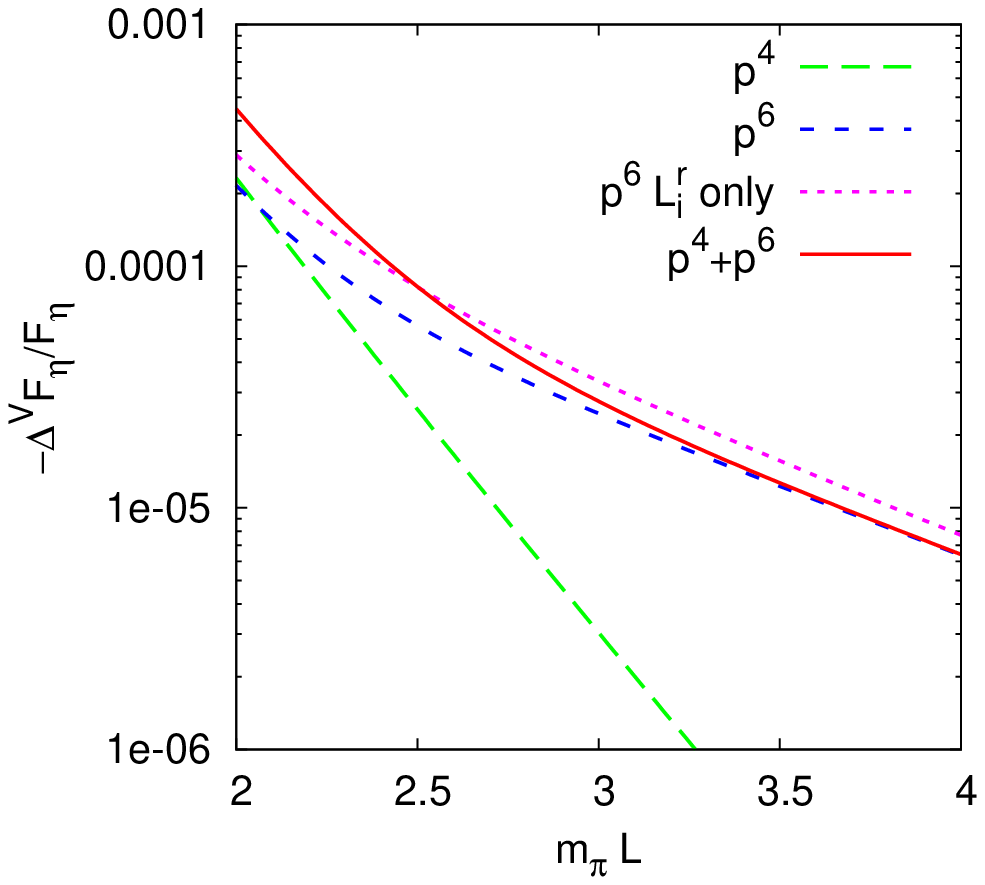}
\centerline{(b)}
\end{minipage}
\caption{The corrections to the kaon and eta decay constant for the physical case.
Plotted is the quantity $-(F_i^{V}-F_i)/F_i$ for $i=K,\eta$.
Shown are the one-loop, the two-loop, the sum and the two-loop $L_i^r$
dependent part.
 (a) Kaon. (b) Eta.}
\label{figfketa}
\end{figure}

The equivalent results for the kaon and eta are plotted in Fig.~\ref{figfketa}.
The kaon decay constant corrections are somewhat smaller than for the pion,
but still important for precision studies.
The one-loop result for the eta decay constant has only a kaon loop
as can be seen from (\ref{minf3p4}). As a result,
that part is very small. The total result comes mainly from
two-loop order. The eta mass has a negative one-loop finite volume
contribution. The pure loop part and the $L_i^r$-dependent part of the $p^6$
contribution are of the expected size. However, there is a very strong
cancellation between the two parts leaving a very small positive correction.
The total finite volume correction for the eta decay constant is quite small.

\begin{figure}[tb!]
\begin{minipage}{0.49\textwidth}
\includegraphics[width=0.99\textwidth]{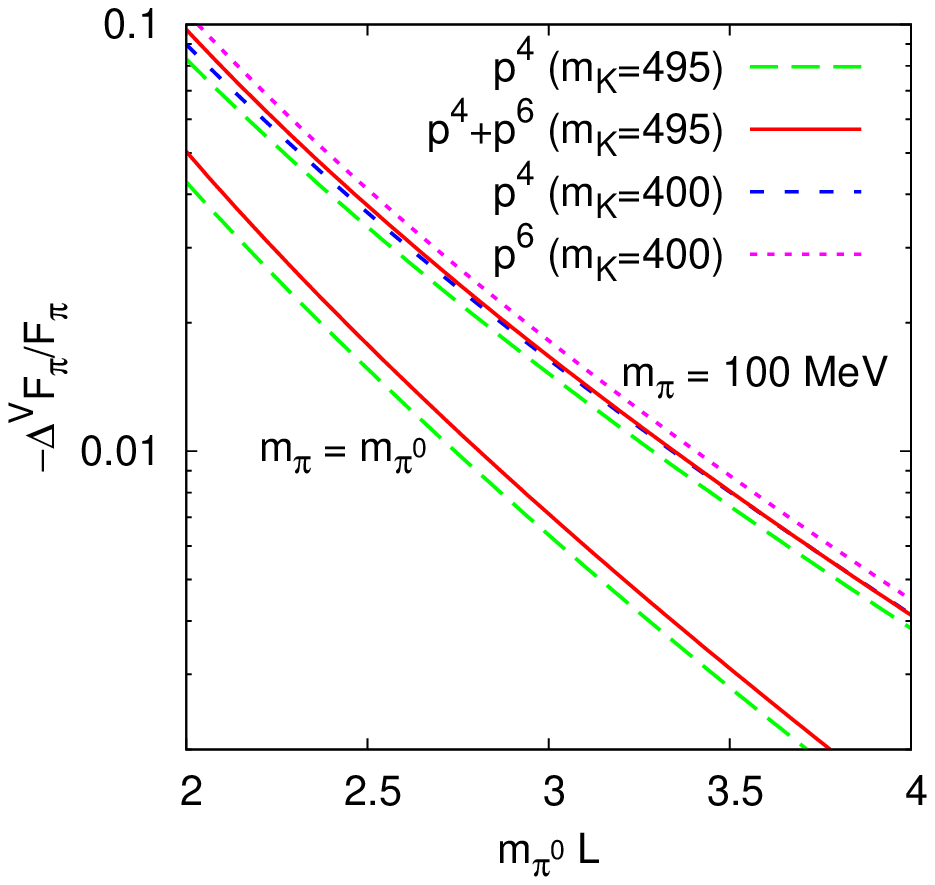}
\centerline{(a)}
\end{minipage}
\begin{minipage}{0.49\textwidth}
\includegraphics[width=0.99\textwidth]{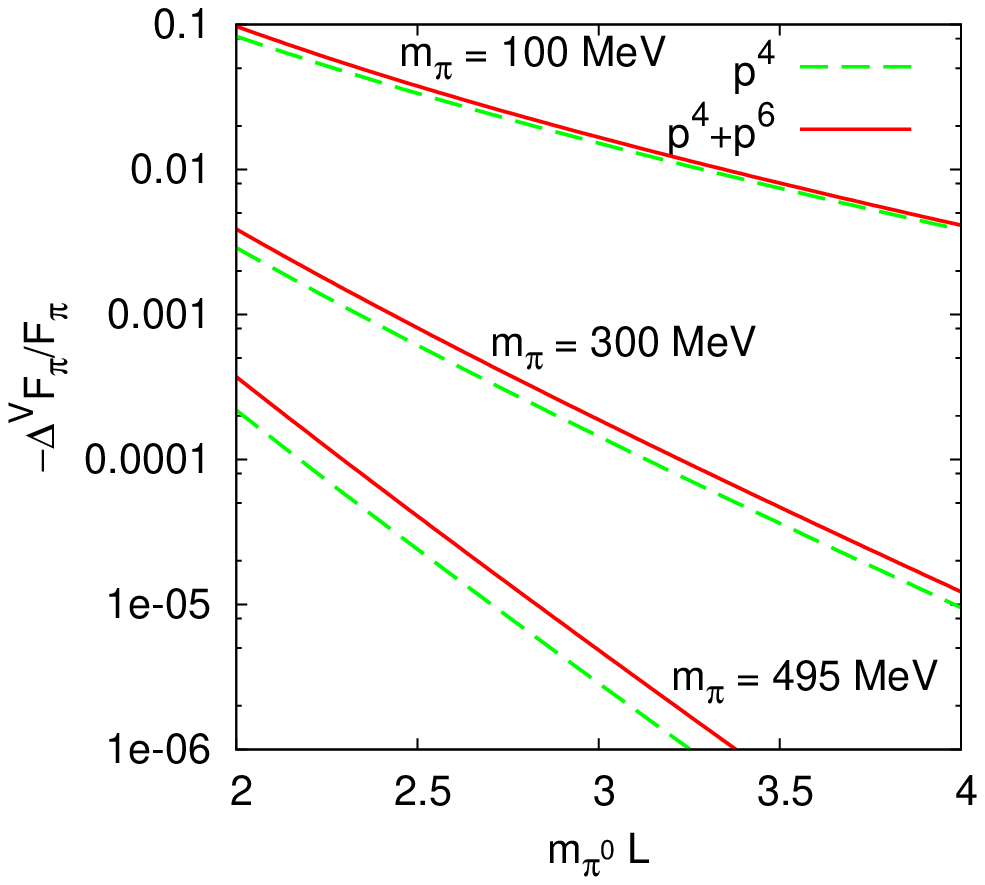}
\centerline{(b)}
\end{minipage}
\caption{The finite volume corrections to the pion decay constant
for a number of cases listed in Tab.~\ref{tabmpimk}.
Plotted is the quantity $-(F_\pi^{V}-F_\pi)/F_\pi$.
(a) Physical case and $(m_\pi,m_K) = (100,495)$ and $(100,400)$~MeV.
(b) $m_K=495$~MeV and $m_\pi=100,300,495$~MeV.
The size $L$ is given in units of the physical $\pi^0$ mass.}
\label{figfpi2}
\end{figure}

We can also check how the finite volume correction depends on the different
masses. In Fig.~\ref{figmpi2} we have plotted the corrections to the
pion decay constant for several scenarios. In Fig.~\ref{figfpi2}(a)
we look at three cases. The bottom two lines are the physical case
labeled with $m_\pi=m_{\pi^0}$ while the top four lines are with
$m_\pi=100$~MeV. There we have plotted two cases, $m_K=400$ and $495$~MeV.
The effect of the change in the pion mass is quite large while the
effect due to the kaon mass change is smaller.
In Fig.~\ref{figfpi2}(b) we can see the effect of only varying the pion mass.

\begin{figure}[tb!]
\begin{minipage}{0.49\textwidth}
\includegraphics[width=0.99\textwidth]{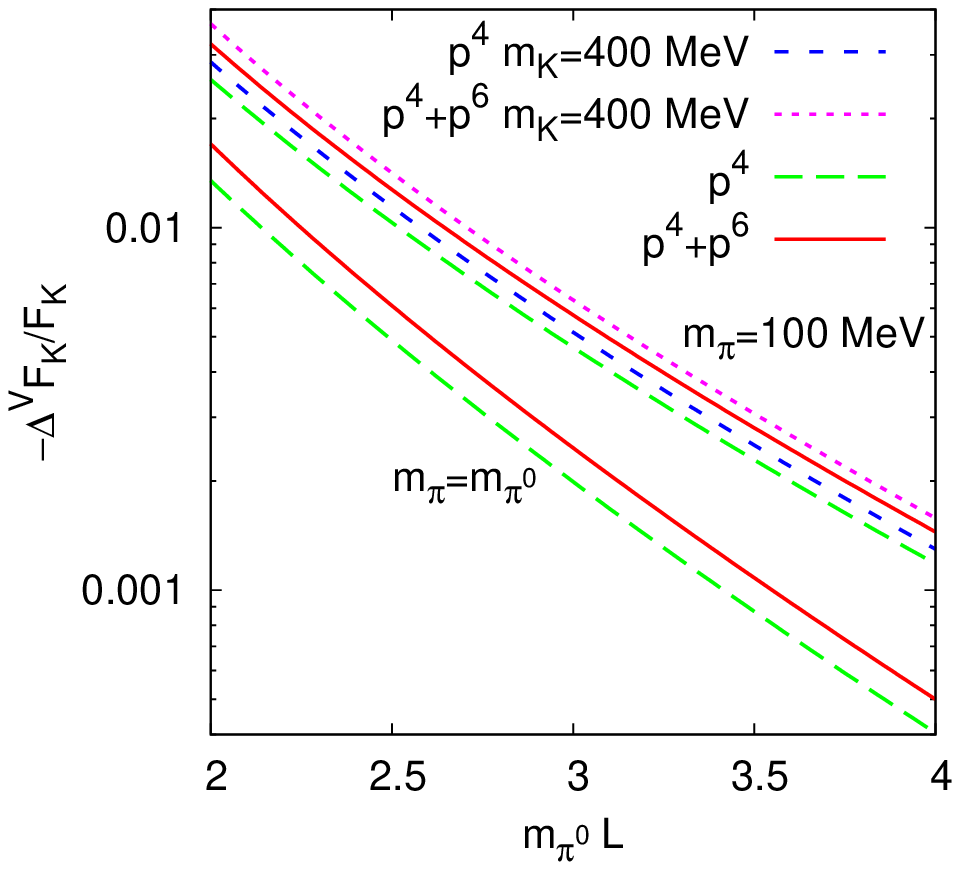}
\centerline{(a)}
\end{minipage}
\begin{minipage}{0.49\textwidth}
\includegraphics[width=0.99\textwidth]{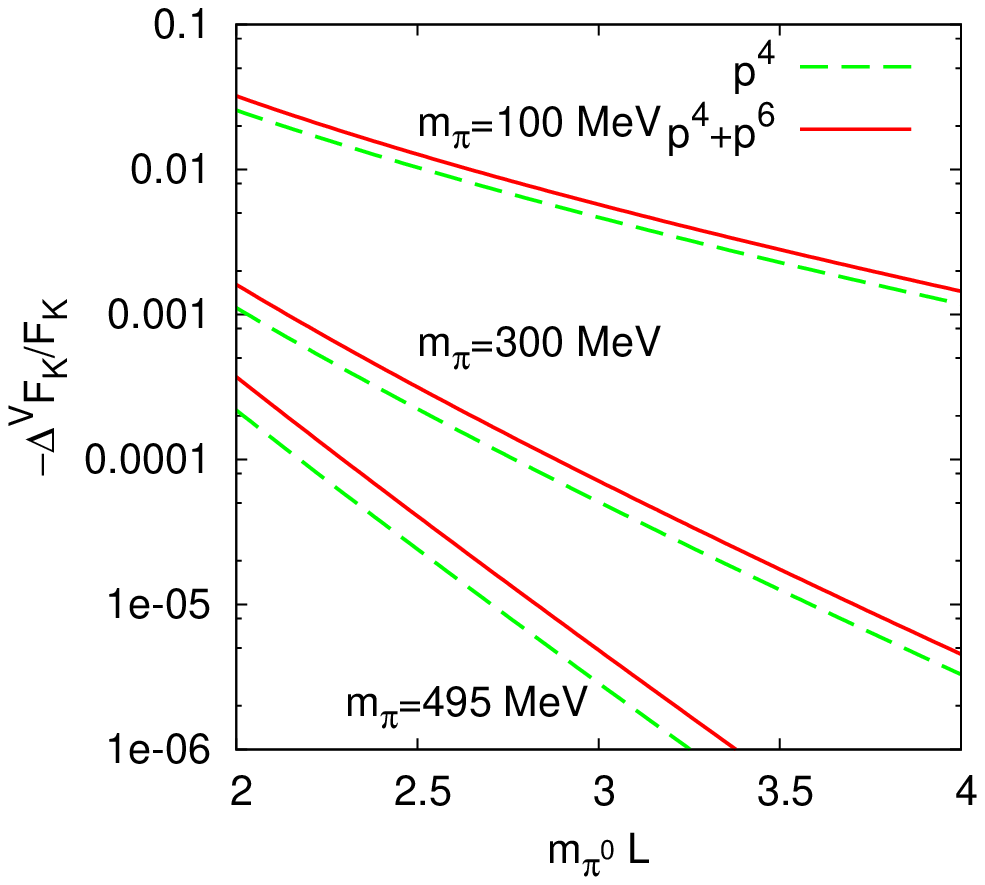}
\centerline{(b)}
\end{minipage}
\caption{The finite volume corrections to the kaon decay constant
for a number of cases listed in Tab.~\ref{tabmpimk}.
Plotted is the quantity $-(F_K^{V}-F_K)/F_K$.
(a) Physical case and $(m_\pi,m_K) = (100,495)$ and $(100,400)$~MeV.
(b) $m_K=495$~MeV and $m_\pi=100,300,495$~MeV.
The size $L$ is given in units of the physical $\pi^0$ mass.}
\label{figfk2}
\end{figure}

We have plotted the same cases for the finite volume corrections to the kaon
decay constant in Fig.~\ref{figfk2}.
In Fig.~\ref{figfk2}(a), the bottom two-lines are the physical case.
The four top lines are with $m_\pi=100$~MeV, where the smaller kaon mass
gives a somewhat larger correction.
In Fig.~\ref{figfk2}(b) we have
shown the corrections for a fixed kaon mass but three different pion masses.
The bottom three lines are the one-loop result while the top three
lines are the full result.
Note that, as it should be, for the case where the pion mass and kaon mass
are the same,
the finite volume corrections to the kaon are the same as for the pion
in Fig.~\ref{figfpi2}(b). This is another small check on our result.

\begin{figure}[tb!]
\begin{minipage}{0.49\textwidth}
\includegraphics[width=0.99\textwidth]{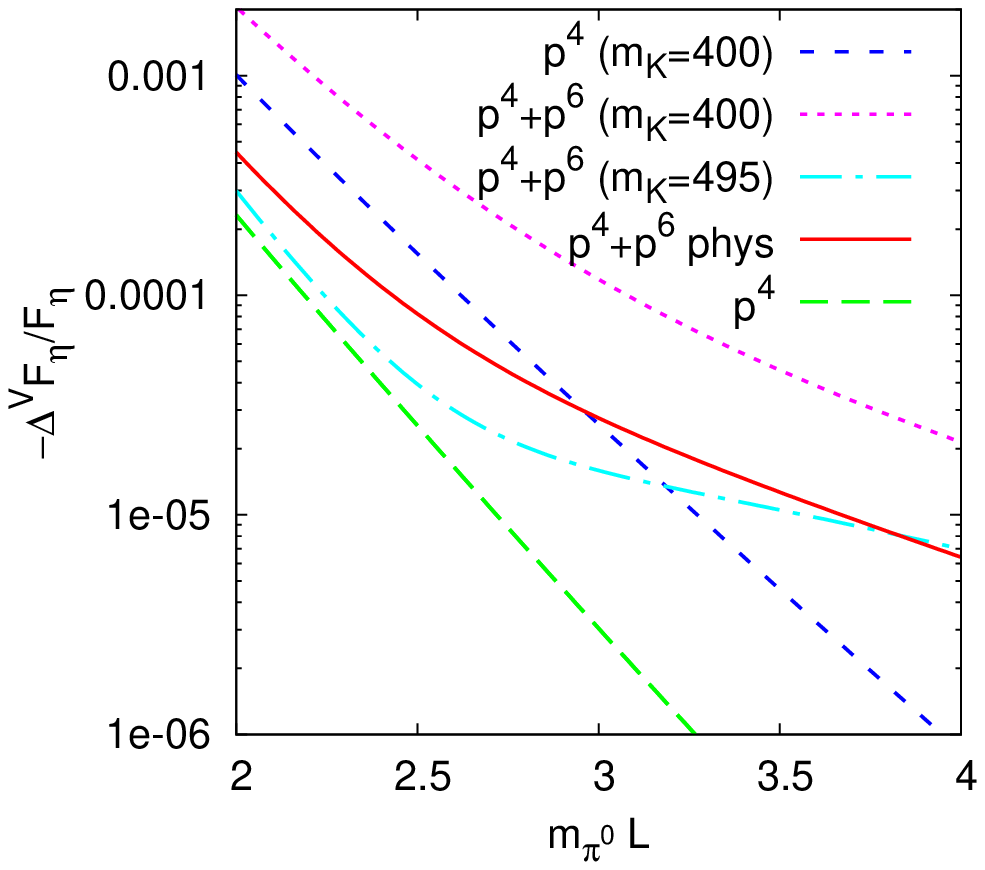}
\centerline{(a)}
\end{minipage}
\begin{minipage}{0.49\textwidth}
\includegraphics[width=0.99\textwidth]{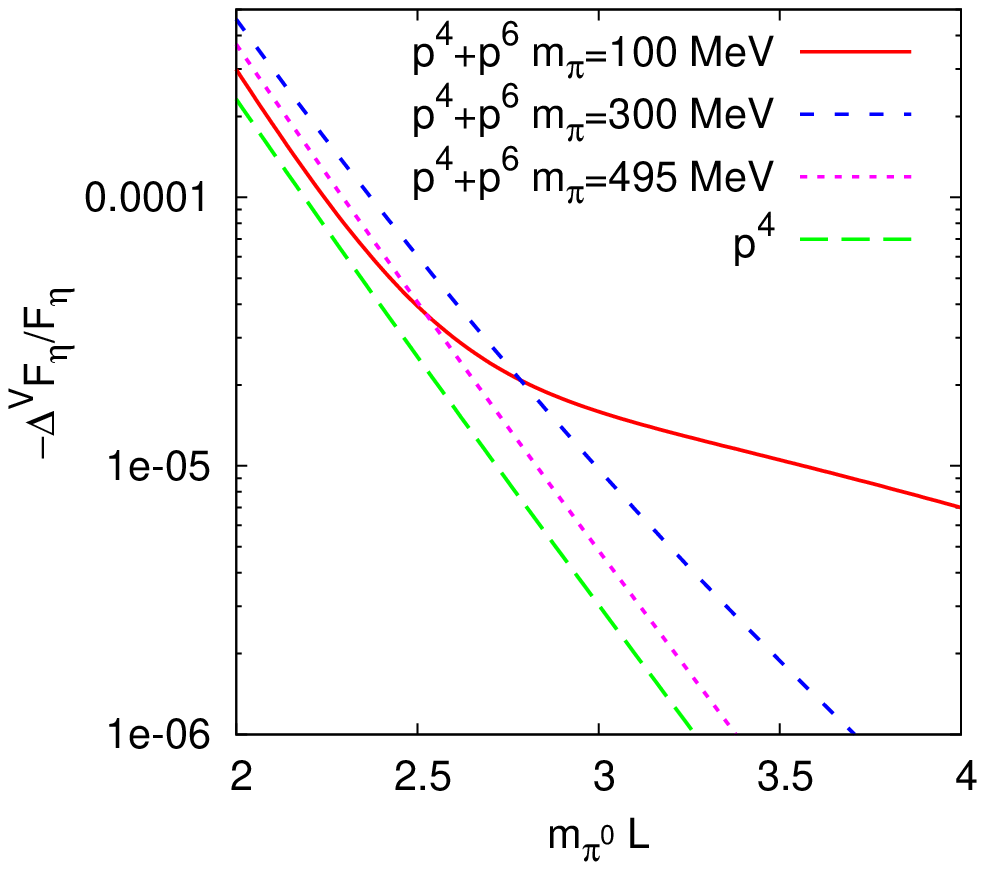}
\centerline{(b)}
\end{minipage}
\caption{The finite volume corrections to the eta decay constant
for a number of cases listed in Tab.~\ref{tabmpimk}.
Plotted is the quantity $-(F_\eta^{V}-F_\eta)/F_\eta$.
(a) Physical case and $(m_\pi,m_K) = (100,495)$ and $(100,400)$~MeV.
The bottom line is the one-loop result for the  physical case and
$(m_\pi,m_K) = (100,495)$. Others as labeled.
(b) $m_K=495$~MeV and $m_\pi=100,300,495$~MeV.
The size $L$ is given in units of the physical $\pi^0$ mass.}
\label{figfeta2}
\end{figure}

We have plotted the same cases once more
for the finite volume corrections to the eta
decay constant squared in Fig.~\ref{figfeta2}.
In Fig.~\ref{figfeta2}(a) the one-loop corrections 
for the physical case and $m_\pi,m_K=100,495$~MeV are extremely close, since
it only depends on the kaon mass.
The $p^6$ corrections for both cases are quite different though.
Finally, for $m_\pi,m_K=100,400$~MeV both the one- and two-loop corrections
are larger but the total correction remains fairly small.
In Fig.~\ref{figfeta2}(b) we have
shown the corrections for a fixed kaon mass but three different pion masses.
The $p^4$ correction is thus identical for the three cases.
The correction for $m_\pi,m_K=495$~MeV agrees with the pion and kaon
corrections for this case.
The total correction remains small for all cases.

We did not compare with the numerical results in \cite{Colangelo:2005gd},
since there was a small mistake in the relevant figures \cite{colangeloprivate}.

\section{Conclusions}

In this paper we calculated the finite volume corrections to two-loop order
in ChPT. The pion mass and decay constant we calculated both in two and
three-flavour ChPT. The kaon and eta mass and decay constant we obtained
in three-flavour ChPT. These expressions in the main text and the appendices
are the main result of this work.

We have compared as far as possible with existing work, where we are in
agreement with the known one-loop results and have some disagreements with the
existing results at two-loop order. What we agree on and differ on
is discussed in Sects.~\ref{sec:nf2} and \ref{sec:nf3}. Note that a full
comparison at the analytical level was not possible due to the large differences
in the loop integral treatments.

We have presented numerical results for a number of representative
cases. In all cases the exponential decay $e^{-m_\pi/L}$ is clearly visible
and as expected the numbers are dominated by the finite volume pion loops.
The corrections at order $p^6$ are sometimes large, especially when
the order $p^4$ result did not contain pion loops. We find that the finite
volume corrections are necessary for the pion mass and decay constant as well
as the kaon decay constant. The kaon mass receives corrections at a somewhat
lower level while finite volume corrections for the eta mass and decay constant
are at present negligible.

The numerical work has been done using {\sc C++}. The programs will be
made available together with the infinite volume results in \cite{chiron}.
The analytical work relied heavily on {\sc FORM} \cite{FORM}.

\section*{Acknowledgements}

We thank Gilberto Colangelo for discussions.
This work is supported in part by the European Community-Research
Infrastructure Integrating Activity ``Study of Strongly Interacting Matter''
(HadronPhysics3, Grant Agreement No. 283286)
and the Swedish Research Council grants 621-2011-5080 and 621-2013-4287.

\appendix
\section{Three flavour $p^6$ expressions for the masses}
\label{appmass}

This appendix lists the order $p^6$ result for the three-flavour ChPT
finite volume corrections to the masses squared at order $p^6$.
\ba
\lefteqn{
F_\pi^4 \Delta^V\! m^{2(6)}_\pi =
        {A}^{V}(m_\pi^2)m_\pi^4 \, \Big( 40\,L^r_{8} + 80\,L^r_{6} - 24\,L^r_{5} - 48\,L^r_{4}
          + 28\,L^r_{3} + 32\,L^r_{2} + 56\,L^r_{1} \Big)
 }
\nonumber\\&&
       + {A}^{V}(m_K^2)m_\pi^2\,m_K^2 \, \Big( 32\,L^r_{8} + 64\,L^r_{6} - 16\,L^r_{5} -
         64\,L^r_{4} + 20\,L^r_{3} + 16\,L^r_{2} + 64\,L^r_{1} \Big)
\nonumber\\&&
       + {A}^{V}(m_\eta^2)m_\pi^2 \, \Big( 8\,L^r_{8}m_\pi^2 - 64/3\,L^r_{7}\,m_K^2 + 64/3\,L^r_{7}m_\pi^2 +
         64/3\,L^r_{6}\,m_K^2 - 16/3\,L^r_{6}m_\pi^2
\nonumber\\&&
 - 32/9\,L^r_{5}m_K^2 + 8/9\,L^r_{5}m_\pi^2
          - 64/3\,L^r_{4}m_K^2 + 16/3\,L^r_{4}m_\pi^2 + 16/3\,L^r_{3}m_K^2 - 4/3\,L^r_{3}
\nonumber\\&&
         m_\pi^2 + 16/3\,L^r_{2}m_K^2 - 4/3\,L^r_{2}m_\pi^2 + 64/3\,L^r_{1}m_K^2 - 16/3\,
         L^r_{1}m_\pi^2 \Big)
\nonumber\\&&
       + {A}_{23}^{V}(m_\pi^2)m_\pi^2 \, \Big(  - 12\,L^r_{3} - 48\,L^r_{2} - 24\,L^r_{1} \Big)
       + {A}_{23}^{V}(m_K^2)m_\pi^2 \, \Big(  - 12\,L^r_{3} - 48\,L^r_{2} \Big)
\nonumber\\&&
       + {A}_{23}^{V}(m_\eta^2)m_\pi^2 \, \Big(  - 4\,L^r_{3} - 12\,L^r_{2}\Big)
\nonumber\\&&
       + {A}^{V}(m_\pi^2)m_\pi^2 \, \Big( \frac{1}{16\pi^2}\,m_K^2 + 3/4\,\frac{1}{16\pi^2}\,m_\pi^2 - 7/4\,\overline{A}(m_\pi^2) -
         \overline{A}(m_K^2) + 1/12\,\overline{B}^0(m_\eta^2)\,m_\pi^2 \Big)
\nonumber\\&&
       + {A}^{V}(m_\pi^2)^2 \, \Big(  - 3/8\,m_\pi^2 \Big)
       + {A}^{V}(m_\pi^2)\,{A}^{V}(m_K^2) \, \Big(  - 1/2\,m_\pi^2 \Big)
\nonumber\\&&
       + {A}^{V}(m_\pi^2)\,{A}^{V}(m_\eta^2) \, \Big(  - 1/12\,m_\pi^2 \Big)
       + {A}^{V}(m_\pi^2)\,{B}^{0V}(m_\pi^2) \, \Big( 1/4\,m_\pi^4 \Big)
\nonumber\\&&
       + {A}^{V}(m_\pi^2)\,{B}^{0V}(m_\eta^2) \, \Big( 1/12\,m_\pi^4 \Big)
\nonumber\\&&
       + {A}^{V}(m_K^2)m_\pi^2 \, \Big( \frac{1}{16\pi^2}m_K^2 - 1/2\,\overline{A}(m_\pi^2) - 1/2\,\overline{A}(m_K^2) -
         1/2\,\overline{A}(m_\eta^2) - 2/9\,\overline{B}^0(m_\eta^2)\,m_K^2 \Big)
\nonumber\\&&
       + {A}^{V}(m_K^2)^2 \, \Big(  - 1/4\,m_\pi^2 \Big)
       + {A}^{V}(m_K^2)\,{A}^{V}(m_\eta^2) \, \Big(  - 1/2\,m_\pi^2 \Big)
       + {A}^{V}(m_K^2)\,{B}^{0V}(m_\eta^2) \, \Big(  - 2/9\,m_\pi^2\,m_K^2 \Big)
\nonumber\\&&
       + {A}^{V}(m_\eta^2)m_\pi^2 \, \Big( 1/12\,\frac{1}{16\pi^2}\,m_\pi^2 + 1/4\,\overline{A}(m_\pi^2) - 1/3\,\overline{A}(m_K^2)
          + \,\overline{B}^0(m_\eta^2)\left(4/27\,m_K^2 - 7/108\,m_\pi^2\right) \Big)
\nonumber\\&&
       + {A}^{V}(m_\eta^2)^2 \, \Big( 1/72\,m_\pi^2 \Big)
       + {A}^{V}(m_\eta^2)\,{B}^{0V}(m_\pi^2) \, \Big(  - 1/12\,m_\pi^4 \Big)
\nonumber\\&&
       + {A}^{V}(m_\eta^2)\,{B}^{0V}(m_\eta^2) \, \Big( 4/27\,m_\pi^2\,m_K^2 - 7/108\,m_\pi^4 \Big)
\nonumber\\&&
       + {H}^V(m_\pi^2,m_\pi^2,m_\pi^2,m_\pi^2) \, \Big( 5/6\,m_\pi^4 \Big)
       + {H}^V(m_\pi^2,m_K^2,m_K^2,m_\pi^2) \, \Big( m_\pi^2\,m_K^2 - 5/8\,m_\pi^4 \Big)
\nonumber\\&&
       + {H}^V(m_\pi^2,m_\eta^2,m_\eta^2,m_\pi^2) \, \Big( 1/18\,m_\pi^4 \Big)
       + {H}^V(m_K^2,m_K^2,m_\eta^2,m_\pi^2) \, \Big( 1/2\,m_\pi^2\,m_K^2 + 1/24\,m_\pi^4 \Big)
\nonumber\\&&
       + {H}_1^V(m_\pi^2,m_K^2,m_K^2,m_\pi^2) \, \Big( m_\pi^4 \Big)
       + {H}_1^V(m_\eta^2,m_K^2,m_K^2,m_\pi^2) \, \Big(  - m_\pi^4 \Big)
\nonumber\\&&
       + {H}_{21}^V(m_\pi^2,m_\pi^2,m_\pi^2,m_\pi^2) \, \Big( 3\,m_\pi^4 \Big)
       + {H}_{21}^V(m_\pi^2,m_K^2,m_K^2,m_\pi^2) \, \Big(  - 3/8\,m_\pi^4 \Big)
\nonumber\\&&
       + {H}_{21}^V(m_K^2,m_\pi^2,m_K^2,m_\pi^2) \, \Big( 3\,m_\pi^4 \Big)
       + {H}_{21}^V(m_\eta^2,m_K^2,m_K^2,m_\pi^2) \, \Big( 9/8\,m_\pi^4 \Big)
\nonumber\\&&
       + {H}_{27}^V(m_\pi^2,m_\pi^2,m_\pi^2,m_\pi^2) \, \Big(  - 3\,m_\pi^2 \Big)
       + {H}_{27}^V(m_\pi^2,m_K^2,m_K^2,m_\pi^2) \, \Big( 3/8\,m_\pi^2 \Big)
\nonumber\\&&
       + {H}_{27}^V(m_K^2,m_\pi^2,m_K^2,m_\pi^2) \, \Big(  - 3\,m_\pi^2 \Big)
       + {H}_{27}^V(m_\eta^2,m_K^2,m_K^2,m_\pi^2) \, \Big(  - 9/8\,m_\pi^2 \Big)
\,.
\ea

\ba
\lefteqn{F_\pi^4 \Delta^V\! m^{2(6)}_K =
        {A}^{V}(m_\pi^2) \, \Big( 24\,L^r_{8}m_\pi^2\,m_K^2 + 48\,L^r_{6}m_\pi^2\,m_K^2 - 12\,L^r_{5}m_\pi^2\,m_K^2 -
         48\,L^r_{4}m_\pi^2\,m_K^2 }
\nonumber\\&&
+ 15\,L^r_{3}m_\pi^2\,m_K^2 + 12\,L^r_{2}m_\pi^2\,m_K^2 + 48\,L^r_{1}m_\pi^2\,m_K^2 \Big)
\nonumber\\&&
       + {A}^{V}(m_K^2) \, \Big( 48\,L^r_{8}m_K^4 + 96\,L^r_{6}m_K^4 - 24\,L^r_{5}m_K^4 - 64\,L^r_{4}
         \,m_K^4 + 30\,L^r_{3}m_K^4 + 36\,L^r_{2}m_K^4 + 72\,L^r_{1}m_K^4 \Big)
\nonumber\\&&
       + {A}^{V}(m_\eta^2) \, \Big( 64/3\,L^r_{8}m_K^4 - 56/3\,L^r_{8}m_\pi^2\,m_K^2 + 16/3\,L^r_{8}m_\pi^4
          + 64/3\,L^r_{7}m_K^4 - 32\,L^r_{7}m_\pi^2\,m_K^2
\nonumber\\&&
 + 32/3\,L^r_{7}m_\pi^4
 + 64/3\,L^r_{6}m_K^4
          - 16/3\,L^r_{6}m_\pi^2\,m_K^2 - 64/9\,L^r_{5}m_K^4 + 4/3\,L^r_{5}m_\pi^2\,m_K^2 - 8/9\,L^r_{5}
         m_\pi^4
\nonumber\\&&
 - 64/3\,L^r_{4}m_K^4 + 16/3\,L^r_{4}m_\pi^2\,m_K^2 + 28/3\,L^r_{3}m_K^4 - 7/3\,L^r_{3}
         m_\pi^2\,m_K^2 + 16/3\,L^r_{2}m_K^4
\nonumber\\&&
 - 4/3\,L^r_{2}m_\pi^2\,m_K^2 + 64/3\,L^r_{1}m_K^4 - 16/3\,
         L^r_{1}m_\pi^2\,m_K^2 \Big)
\nonumber\\&&
       + {A}_{23}^{V}(m_\pi^2) \, \Big(  - 9\,L^r_{3}m_K^2 - 36\,L^r_{2}m_K^2 \Big)
       + {A}_{23}^{V}(m_K^2) \, \Big(  - 18\,L^r_{3}m_K^2 - 60\,L^r_{2}m_K^2 - 24\,L^r_{1}m_K^2 \Big)
\nonumber\\&&
       + {A}_{23}^{V}(m_\eta^2) \, \Big(  - L^r_{3}m_K^2 - 12\,L^r_{2}m_K^2 \Big)
\nonumber\\&&
       + {A}^{V}(m_\pi^2) \, \Big( 3/4\,\frac{1}{16\pi^2}\,m_K^4 - 3/16\,\overline{A}(m_\pi^2)\,m_K^2 - 3/4\,\overline{A}(m_K^2)\,m_K^2
          - 3/16\,\overline{A}(m_\eta^2)\,m_K^2
\nonumber\\&&
 - 1/6\,\overline{B}^0(m_\eta^2)\,m_\pi^2\,m_K^2 \Big)
\nonumber\\&&
       + {A}^{V}(m_\pi^2)^2 \, \Big(  - 3/32\,m_K^2 \Big)
       + {A}^{V}(m_\pi^2)\,{A}^{V}(m_K^2) \, \Big(  - 3/4\,m_K^2 \Big)
       + {A}^{V}(m_\pi^2)\,{A}^{V}(m_\eta^2) \, \Big(  - 3/16\,m_K^2 \Big)
\nonumber\\&&
       + {A}^{V}(m_\pi^2)\,{B}^{0V}(m_\eta^2) \, \Big(  - 1/6\,m_\pi^2\,m_K^2 \Big)
\nonumber\\&&
       + {A}^{V}(m_K^2) \, \Big( 3/4\,\frac{1}{16\pi^2}\,m_K^4 + 3/4\,\frac{1}{16\pi^2}\,m_\pi^2\,m_K^2 - 3/4\,\overline{A}(m_\pi^2)\,m_K^2
          - 3/2\,\overline{A}(m_K^2)\,m_K^2
\nonumber\\&&
 + 4/9\,\overline{B}^0(m_\eta^2)\,m_K^4 \Big)
\nonumber\\&&
       + {A}^{V}(m_K^2)^2 \, \Big(  - 3/4\,m_K^2 \Big)
       + {A}^{V}(m_K^2)\,{B}^{0V}(m_\eta^2) \, \Big( 4/9\,m_K^4 \Big)
\nonumber\\&&
       + {A}^{V}(m_\eta^2) \, \Big( 1/2\,\frac{1}{16\pi^2}\,m_K^4 + 1/4\,\frac{1}{16\pi^2}\,m_\pi^2\,m_K^2 - 41/48\,\overline{A}(m_\pi^2)\,
         m_K^2 + 1/12\,\overline{A}(m_\pi^2)\,m_\pi^2 
\nonumber\\&&
- 2/3\,\overline{A}(m_K^2)\,m_K^2
 + 19/48\,\overline{A}(m_\eta^2)\,m_K^2 - 1/12\,
         \overline{A}(m_\eta^2)\,m_\pi^2 - 8/27\,\overline{B}^0(m_\eta^2)\,m_K^4
\nonumber\\&&
 + 7/54\,\overline{B}^0(m_\eta^2)\,m_\pi^2\,m_K^2 \Big)
\nonumber\\&&
       + {A}^{V}(m_\eta^2)^2 \, \Big( 25/288\,m_K^2 \Big)
       + {A}^{V}(m_\eta^2)\,{B}^{0V}(m_\eta^2) \, \Big(  - 8/27\,m_K^4 + 7/54\,m_\pi^2\,m_K^2 \Big)
\nonumber\\&&
       + {H}^V(m_\pi^2,m_\pi^2,m_K^2,m_K^2) \, \Big(  - 15/32\,m_K^4 + 3/4\,m_\pi^2\,m_K^2 \Big)
\nonumber\\&&
       + {H}^V(m_\pi^2,m_K^2,m_\eta^2,m_K^2) \, \Big( 13/16\,m_K^4 \Big)
       + {H}^V(m_K^2,m_K^2,m_K^2,m_K^2) \, \Big( 3/4\,m_K^4 \Big)
\nonumber\\&&
       + {H}^V(m_K^2,m_\eta^2,m_\eta^2,m_K^2) \, \Big( 181/288\,m_K^4 \Big)
       + {H}_1^V(m_K^2,m_\pi^2,m_\pi^2,m_K^2) \, \Big( 3/4\,m_K^4 \Big)
\nonumber\\&&
       + {H}_1^V(m_K^2,m_\pi^2,m_\eta^2,m_K^2) \, \Big(  - 3/2\,m_K^4 \Big)
       + {H}_1^V(m_K^2,m_\eta^2,m_\eta^2,m_K^2) \, \Big(  - 5/4\,m_K^4 \Big)
\nonumber\\&&
       + {H}_{21}^V(m_\pi^2,m_\pi^2,m_K^2,m_K^2) \, \Big( 9/4\,m_K^4 \Big)
       + {H}_{21}^V(m_K^2,m_\pi^2,m_\pi^2,m_K^2) \, \Big(  - 9/32\,m_K^4 \Big)
\nonumber\\&&
       + {H}_{21}^V(m_K^2,m_\pi^2,m_\eta^2,m_K^2) \, \Big( 27/16\,m_K^4 \Big)
       + {H}_{21}^V(m_K^2,m_K^2,m_K^2,m_K^2) \, \Big( 9/4\,m_K^4 \Big)
\nonumber\\&&
       + {H}_{21}^V(m_K^2,m_\eta^2,m_\eta^2,m_K^2) \, \Big( 27/32\,m_K^4 \Big)
       + {H}_{27}^V(m_\pi^2,m_\pi^2,m_K^2,m_K^2) \, \Big(  - 9/4\,m_K^2 \Big)
\nonumber\\&&
       + {H}_{27}^V(m_K^2,m_\pi^2,m_\pi^2,m_K^2) \, \Big( 9/32\,m_K^2 \Big)
       + {H}_{27}^V(m_K^2,m_\pi^2,m_\eta^2,m_K^2) \, \Big(  - 27/16\,m_K^2 \Big)
\nonumber\\&&
       + {H}_{27}^V(m_K^2,m_K^2,m_K^2,m_K^2) \, \Big(  - 9/4\,m_K^2 \Big)
       + {H}_{27}^V(m_K^2,m_\eta^2,m_\eta^2,m_K^2) \, \Big(  - 27/32\,m_K^2 \Big)\,.
\ea

\ba
\lefteqn{F_\pi^4 \Delta^V\! m^{2(6)}_\eta =
        {A}^{V}(m_\pi^2) \, \Big( 24\,L^r_{8}m_\pi^4 - 64\,L^r_{7}m_\pi^2\,m_K^2 + 64\,L^r_{7}m_\pi^4 + 64\,
         L^r_{6}m_\pi^2\,m_K^2
 - 16\,L^r_{6}m_\pi^4}
\nonumber\\&&
 - 32/3\,L^r_{5}m_\pi^2\,m_K^2
 + 8/3\,L^r_{5}m_\pi^4 - 64\,
         L^r_{4}m_\pi^2\,m_K^2 
+ 16\,L^r_{4}m_\pi^4 + 16\,L^r_{3}m_\pi^2\,m_K^2 - 4\,L^r_{3}m_\pi^4 
\nonumber\\&&
+ 16\,L^r_{2}
         m_\pi^2\,m_K^2 - 4\,L^r_{2}m_\pi^4 + 64\,L^r_{1}m_\pi^2\,m_K^2 - 16\,L^r_{1}m_\pi^4 \Big)
\nonumber\\&&
       + {A}^{V}(m_K^2) \, \Big( 256/3\,L^r_{8}m_K^4 - 224/3\,L^r_{8}m_\pi^2\,m_K^2 + 64/3\,L^r_{8}
         m_\pi^4 + 256/3\,L^r_{7}m_K^4
\nonumber\\&&
 - 128\,L^r_{7}m_\pi^2\,m_K^2 + 128/3\,L^r_{7}m_\pi^4 + 256/3\,
         L^r_{6}m_K^4 - 64/3\,L^r_{6}m_\pi^2\,m_K^2 - 256/9\,L^r_{5}m_K^4
\nonumber\\&&
 + 16/3\,L^r_{5}m_\pi^2\,m_K^2 -
         32/9\,L^r_{5}m_\pi^4 - 256/3\,L^r_{4}m_K^4 + 64/3\,L^r_{4}m_\pi^2\,m_K^2 + 112/3\,L^r_{3}m_K^4
\nonumber\\&&
          - 28/3\,L^r_{3}m_\pi^2\,m_K^2 + 64/3\,L^r_{2}m_K^4 - 16/3\,L^r_{2}m_\pi^2\,m_K^2 + 256/3\,L^r_{1}
         m_K^4 - 64/3\,L^r_{1}m_\pi^2\,m_K^2 \Big)
\nonumber\\&&
       + {A}^{V}(m_\eta^2) \, \Big( 896/9\,L^r_{8}m_K^4 - 1024/9\,L^r_{8}m_\pi^2\,m_K^2 + 344/9\,L^r_{8}
         m_\pi^4 + 1024/9\,L^r_{7}m_K^4
\nonumber\\&&
 - 1664/9\,L^r_{7}m_\pi^2\,m_K^2 + 640/9\,L^r_{7}m_\pi^4 + 256/
         3\,L^r_{6}m_K^4 - 128/3\,L^r_{6}m_\pi^2\,m_K^2 + 16/3\,L^r_{6}m_\pi^4
\nonumber\\&&
 - 832/27\,L^r_{5}m_K^4
          + 896/27\,L^r_{5}m_\pi^2\,m_K^2 - 280/27\,L^r_{5}m_\pi^4 - 256/9\,L^r_{4}m_K^4 + 128/9\,
         L^r_{4}m_\pi^2\,m_K^2
\nonumber\\&&
 - 16/9\,L^r_{4}m_\pi^4 + 64/3\,L^r_{3}m_K^4 - 32/3\,L^r_{3}m_\pi^2\,m_K^2 + 4/
         3\,L^r_{3}m_\pi^4 + 128/3\,L^r_{2}m_K^4
\nonumber\\&&
 - 64/3\,L^r_{2}m_\pi^2\,m_K^2 + 8/3\,L^r_{2}m_\pi^4 +
         128/3\,L^r_{1}m_K^4 - 64/3\,L^r_{1}m_\pi^2\,m_K^2 + 8/3\,L^r_{1}m_\pi^4 \Big)
\nonumber\\&&
       + {A}_{23}^{V}(m_\pi^2) \, \Big(  - 16\,L^r_{3}m_K^2 + 4\,L^r_{3}m_\pi^2 - 48\,L^r_{2}m_K^2 + 12\,L^r_{2}
         m_\pi^2 \Big)
\nonumber\\&&
       + {A}_{23}^{V}(m_K^2) \, \Big(  - 16/3\,L^r_{3}m_K^2 + 4/3\,L^r_{3}m_\pi^2 - 64\,L^r_{2}m_K^2 + 16\,
         L^r_{2}m_\pi^2 \Big)
\nonumber\\&&
       + {A}_{23}^{V}(m_\eta^2) \, \Big(  - 16\,L^r_{3}m_K^2 + 4\,L^r_{3}m_\pi^2 - 32\,L^r_{2}m_K^2 + 8\,L^r_{2}
         m_\pi^2 - 32\,L^r_{1}m_K^2 + 8\,L^r_{1}m_\pi^2 \Big)
\nonumber\\&&
       + {A}^{V}(m_\pi^2) \, \Big( 1/4\,\frac{1}{16\pi^2}\,m_\pi^4 + 3/4\,\overline{A}(m_\pi^2)\,m_\pi^2 - \overline{A}(m_K^2)\,m_\pi^2 + 4/
         9\,\overline{B}^0(m_\eta^2)\,m_\pi^2\,m_K^2
\nonumber\\&&
 - 7/36\,\overline{B}^0(m_\eta^2)\,m_\pi^4 \Big)
\nonumber\\&&
       + {A}^{V}(m_\pi^2)^2 \, \Big(  - 1/8\,m_\pi^2 \Big)
       + {A}^{V}(m_\pi^2)\,{A}^{V}(m_K^2) \, \Big(  - 3/2\,m_\pi^2 \Big)
       + {A}^{V}(m_\pi^2)\,{A}^{V}(m_\eta^2) \, \Big( 1/12\,m_\pi^2 \Big)
\nonumber\\&&
       + {A}^{V}(m_\pi^2)\,{B}^{0V}(m_\pi^2) \, \Big(  - 1/4\,m_\pi^4 \Big)
       + {A}^{V}(m_\pi^2)\,{B}^{0V}(m_\eta^2) \, \Big( 4/9\,m_\pi^2\,m_K^2 - 7/36\,m_\pi^4 \Big)
\nonumber\\&&
       + {A}^{V}(m_K^2) \, \Big( 8/3\,\frac{1}{16\pi^2}\,m_K^4 + 2/3\,\frac{1}{16\pi^2}\,m_\pi^2\,m_K^2 - 1/3\,\frac{1}{16\pi^2}\,m_\pi^4
          - 8/3\,\overline{A}(m_\pi^2)\,m_K^2
\nonumber\\&&
 - 7/6\,\overline{A}(m_\pi^2)\,m_\pi^2 - 2/3\,\overline{A}(m_K^2)\,m_K^2 + 3/2\,\overline{A}(m_K^2)\,
         m_\pi^2 - 8/3\,\overline{A}(m_\eta^2)\,m_K^2 + 7/6\,\overline{A}(m_\eta^2)\,m_\pi^2
\nonumber\\&&
 - 32/27\,\overline{B}^0(m_\eta^2)\,m_K^4 +
         14/27\,\overline{B}^0(m_\eta^2)\,m_\pi^2\,m_K^2 \Big)
\nonumber\\&&
       + {A}^{V}(m_K^2)^2 \, \Big( m_K^2 + 3/4\,m_\pi^2 \Big)
       + {A}^{V}(m_K^2)\,{A}^{V}(m_\eta^2) \, \Big(  - 32/9\,m_K^2 + 3/2\,m_\pi^2 \Big)
\nonumber\\&&
       + {A}^{V}(m_K^2)\,{B}^{0V}(m_\eta^2) \, \Big(  - 32/27\,m_K^4 + 14/27\,m_\pi^2\,m_K^2 \Big)
\nonumber\\&&
       + {A}^{V}(m_\eta^2) \, \Big(  - 4/9\,\frac{1}{16\pi^2}\,m_K^4 - 1/12\,\frac{1}{16\pi^2}\,m_\pi^4 + 16/9\,\overline{A}(m_\pi^2)\,
         m_K^2 - 29/36\,\overline{A}(m_\pi^2)\,m_\pi^2
\nonumber\\&&
 - 20/9\,\overline{A}(m_K^2)\,m_K^2 + 10/9\,\overline{A}(m_K^2)\,m_\pi^2 + 8/9\,
         \overline{A}(m_\eta^2)\,m_K^2 - 2/9\,\overline{A}(m_\eta^2)\,m_\pi^2
\nonumber\\&&
 + 64/81\,\overline{B}^0(m_\eta^2)\,m_K^4 - 56/81\,\overline{B}^0(
         m_\eta^2)\,m_\pi^2\,m_K^2 + 49/324\,\overline{B}^0(m_\eta^2)\,m_\pi^4 \Big)
\nonumber\\&&
       + {A}^{V}(m_\eta^2)^2 \, \Big( 8/27\,m_K^2 - 31/216\,m_\pi^2 \Big)
       + {A}^{V}(m_\eta^2)\,{B}^{0V}(m_\pi^2) \, \Big( 1/12\,m_\pi^4 \Big)
\nonumber\\&&
       + {A}^{V}(m_\eta^2)\,{B}^{0V}(m_K^2) \, \Big( 4/9\,m_K^4 \Big)
\nonumber\\&&
       + {A}^{V}(m_\eta^2)\,{B}^{0V}(m_\eta^2) \, \Big( 64/81\,m_K^4 - 56/81\,m_\pi^2\,m_K^2 + 49/324\,
         m_\pi^4 \Big)
\nonumber\\&&
       + {H}^V(m_\pi^2,m_\pi^2,m_\eta^2,m_\eta^2) \, \Big( 1/6\,m_\pi^4 \Big)
       + {H}^V(m_\pi^2,m_K^2,m_K^2,m_\eta^2) \, \Big( 3/2\,m_\pi^2\,m_K^2 + 1/8\,m_\pi^4 \Big)
\nonumber\\&&
       + {H}^V(m_K^2,m_K^2,m_\eta^2,m_\eta^2) \, \Big( 50/9\,m_K^4 - 11/3\,m_\pi^2\,m_K^2 + 5/8\,m_\pi^4
          \Big)
\nonumber\\&&
       + {H}^V(m_\eta^2,m_\eta^2,m_\eta^2,m_\eta^2) \, \Big( 128/243\,m_K^4 - 112/243\,m_\pi^2\,m_K^2 + 49/
         486\,m_\pi^4 \Big)
\nonumber\\&&
       + {H}_1^V(m_\pi^2,m_K^2,m_K^2,m_\eta^2) \, \Big(  - 4\,m_\pi^2\,m_K^2 + m_\pi^4 \Big)
\nonumber\\&&
       + {H}_1^V(m_\eta^2,m_K^2,m_K^2,m_\eta^2) \, \Big(  - 32/3\,m_K^4 + 20/3\,m_\pi^2\,m_K^2 - m_\pi^4
          \Big)
\nonumber\\&&
       + {H}_{21}^V(m_\pi^2,m_K^2,m_K^2,m_\eta^2) \, \Big( 6\,m_K^4 - 3\,m_\pi^2\,m_K^2 + 3/8\,m_\pi^4 \Big)
\nonumber\\&&
       + {H}_{21}^V(m_\eta^2,m_K^2,m_K^2,m_\eta^2) \, \Big( 6\,m_K^4 - 3\,m_\pi^2\,m_K^2 + 3/8\,m_\pi^4 \Big)
\nonumber\\&&
       + {H}_{27}^V(m_\pi^2,m_K^2,m_K^2,m_\eta^2) \, \Big(  - 9/2\,m_K^2 + 9/8\,m_\pi^2 \Big)
\nonumber\\&&
       + {H}_{27}^V(m_\eta^2,m_K^2,m_K^2,m_\eta^2) \, \Big(  - 9/2\,m_K^2 + 9/8\,m_\pi^2 \Big)\,.
\ea

\section{Three flavour $p^6$ expressions for the decay constants}
\label{appdecay}

This appendix lists the order $p^6$ result for the three-flavour ChPT
finite volume corrections to the decay constants at order $p^6$.
\ba
\nonumber\\&&
\lefteqn{  F_\pi^3\Delta^VF_\pi^{(6)} =
        {A}^{V}(m_\pi^2) \, \Big( 6\,L^r_{5}m_\pi^2 + 12\,L^r_{4}m_\pi^2 - 14\,L^r_{3}m_\pi^2 - 16\,L^r_{2}m_\pi^2 -
         28\,L^r_{1}m_\pi^2 \Big)}
\nonumber\\&&
       + {A}^{V}(m_K^2) \, \Big( 4\,L^r_{5}m_\pi^2 + 16\,L^r_{4}m_K^2 - 10\,L^r_{3}m_K^2 - 8\,L^r_{2}m_K^2 -
         32\,L^r_{1}m_K^2 \Big)
\nonumber\\&&
       + {A}^{V}(m_\eta^2) \, \Big( 2/3\,L^r_{5}m_\pi^2 + 16/3\,L^r_{4}m_K^2 - 4/3\,L^r_{4}m_\pi^2 - 8/3\,L^r_{3}
         m_K^2 + 2/3\,L^r_{3}m_\pi^2
\nonumber\\&&
 - 8/3\,L^r_{2}m_K^2
 + 2/3\,L^r_{2}m_\pi^2 - 32/3\,L^r_{1}m_K^2 + 8/3\,
         L^r_{1}m_\pi^2 \Big)
\nonumber\\&&
       + {A}_{23}^{V}(m_\pi^2) \, \Big( 6\,L^r_{3}+ 24\,L^r_{2}+ 12\,L^r_{1})
       + {A}_{23}^{V}(m_K^2) \, \Big( 6\,L^r_{3}+ 24\,L^r_{2})
       + {A}_{23}^{V}(m_\eta^2) \, \Big( 2\,L^r_{3}+ 6\,L^r_{2})
\nonumber\\&&
       + {A}^{V}(m_\pi^2) \, \Big(  - 1/2\,\frac{1}{16\pi^2}\,m_K^2 - 1/4\,\frac{1}{16\pi^2}\,m_\pi^2 + 1/2\,\overline{A}(m_\pi^2) + 1/2\,
         \overline{A}(m_K^2) \Big)
\nonumber\\&&
       + {A}^{V}(m_\pi^2)\,{B}^{0V}(m_\pi^2) \, \Big(  - 1/2\,m_\pi^2 \Big)
\nonumber\\&&
       + {A}^{V}(m_K^2) \, \Big(  - 1/2\,\frac{1}{16\pi^2}\,m_K^2 - 1/8\,\frac{1}{16\pi^2}\,m_\pi^2 + 1/2\,\overline{A}(m_\pi^2) + 1/4\,
         \overline{A}(m_K^2) \Big)
\nonumber\\&&
       + {A}^{V}(m_\eta^2) \, \Big( 1/6\,\frac{1}{16\pi^2}\,m_K^2 - 1/6\,\frac{1}{16\pi^2}\,m_\pi^2 + 1/6\,\overline{A}(m_\pi^2) - 1/6\,\overline{A}(
         m_K^2) \Big)
\nonumber\\&&
       + {A}^{V}(m_\eta^2)\,{B}^{0V}(m_\pi^2) \, \Big( 1/6\,m_\pi^2 \Big)
       + {A}^{V}(m_\eta^2)\,{B}^{0V}(m_K^2) \, \Big(  - 1/6\,m_K^2 \Big)
\nonumber\\&&
       + {H}^V(m_\pi^2,m_\pi^2,m_\pi^2,m_\pi^2) \, \Big(  - 1/2\,m_\pi^2 \Big)
       + {H}^V(m_\pi^2,m_K^2,m_K^2,m_\pi^2) \, \Big(  - 1/2\,m_K^2 + 1/16\,m_\pi^2 \Big)
\nonumber\\&&
       + {H}^V(m_K^2,m_K^2,m_\eta^2,m_\pi^2) \, \Big(  - 1/4\,m_K^2 + 1/16\,m_\pi^2 \Big)
       + {H}_{27}^V(m_\pi^2,m_\pi^2,m_\pi^2,m_\pi^2) \, \Big( 3/2 \Big)
\nonumber\\&&
       + {H}_{27}^V(m_\pi^2,m_K^2,m_K^2,m_\pi^2) \, \Big(  - 3/16 \Big)
       + {H}_{27}^V(m_K^2,m_\pi^2,m_K^2,m_\pi^2) \, \Big( 3/2 \Big)
\nonumber\\&&
       + {H}_{27}^V(m_\eta^2,m_K^2,m_K^2,m_\pi^2) \, \Big( 9/16 \Big)
       + {H}^{V\prime}(m_\pi^2,m_\pi^2,m_\pi^2,m_\pi^2) \, \Big( 5/12\,m_\pi^4 \Big)
\nonumber\\&&
       + {H}^{V\prime}(m_\pi^2,m_K^2,m_K^2,m_\pi^2) \, \Big( 1/2\,m_\pi^2\,m_K^2 - 5/16\,m_\pi^4 \Big)
       + {H}^{V\prime}(m_\pi^2,m_\eta^2,m_\eta^2,m_\pi^2) \, \Big( 1/36\,m_\pi^4 \Big)
\nonumber\\&&
       + {H}^{V\prime}(m_K^2,m_K^2,m_\eta^2,m_\pi^2) \, \Big( 1/4\,m_\pi^2\,m_K^2 + 1/48\,m_\pi^4 \Big)
       + {H}_1^{V\prime}(m_\pi^2,m_K^2,m_K^2,m_\pi^2) \, \Big( 1/2\,m_\pi^4 \Big)
\nonumber\\&&
       + {H}_1^{V\prime}(m_\eta^2,m_K^2,m_K^2,m_\pi^2) \, \Big(  - 1/2\,m_\pi^4 \Big)
       + {H}_{21}^{V\prime}(m_\pi^2,m_\pi^2,m_\pi^2,m_\pi^2) \, \Big( 3/2\,m_\pi^4 \Big)
\nonumber\\&&
       + {H}_{21}^{V\prime}(m_\pi^2,m_K^2,m_K^2,m_\pi^2) \, \Big(  - 3/16\,m_\pi^4 \Big)
       + {H}_{21}^{V\prime}(m_K^2,m_\pi^2,m_K^2,m_\pi^2) \, \Big( 3/2\,m_\pi^4 \Big)
\nonumber\\&&
       + {H}_{21}^{V\prime}(m_\eta^2,m_K^2,m_K^2,m_\pi^2) \, \Big( 9/16\,m_\pi^4 \Big)
       + {H}_{27}^{V\prime}(m_\pi^2,m_\pi^2,m_\pi^2,m_\pi^2) \, \Big(  - 3/2\,m_\pi^2 \Big)
\nonumber\\&&
       + {H}_{27}^{V\prime}(m_\pi^2,m_K^2,m_K^2,m_\pi^2) \, \Big( 3/16\,m_\pi^2 \Big)
       + {H}_{27}^{V\prime}(m_K^2,m_\pi^2,m_K^2,m_\pi^2) \, \Big(  - 3/2\,m_\pi^2 \Big)
\nonumber\\&&
       + {H}_{27}^{V\prime}(m_\eta^2,m_K^2,m_K^2,m_\pi^2) \, \Big(  - 9/16\,m_\pi^2 \Big)\,.
\ea

\ba
\lefteqn{F_\pi^3 \Delta^V F_K^{(6)}=
        {A}^{V}(m_\pi^2) \, \Big( 3/2\,L^r_{5}m_K^2 + 3/2\,L^r_{5}m_\pi^2 + 12\,L^r_{4}m_\pi^2 - 15/2\,L^r_{3} 
         m_\pi^2 - 6\,L^r_{2}m_\pi^2 }
\nonumber\\&&
 - 24\,L^r_{1}m_\pi^2 \Big)
\nonumber\\&&
       + {A}^{V}(m_K^2) \, \Big( 3\,L^r_{5}m_K^2 + 3\,L^r_{5}m_\pi^2 + 16\,L^r_{4}m_K^2 - 15\,L^r_{3}m_K^2 -
         18\,L^r_{2}m_K^2 - 36\,L^r_{1}m_K^2 \Big)
\nonumber\\&&
       + {A}^{V}(m_\eta^2) \, \Big( 1/6\,L^r_{5}m_K^2 + 3/2\,L^r_{5}m_\pi^2 + 16/3\,L^r_{4}m_K^2 - 4/3\,L^r_{4}
         m_\pi^2 - 14/3\,L^r_{3}m_K^2
\nonumber\\&&
 + 7/6\,L^r_{3}m_\pi^2 - 8/3\,L^r_{2}m_K^2 + 2/3\,L^r_{2}m_\pi^2 - 32/3\,
         L^r_{1}m_K^2 + 8/3\,L^r_{1}m_\pi^2 \Big)
\nonumber\\&&
       + {A}_{23}^{V}(m_\pi^2) \, \Big( 9/2\,L^r_{3}+ 18\,L^r_{2})
       + {A}_{23}^{V}(m_K^2) \, \Big( 9\,L^r_{3}+ 30\,L^r_{2}+ 12\,L^r_{1})
\nonumber\\&&
       + {A}_{23}^{V}(m_\eta^2) \, \Big( 1/2\,L^r_{3}+ 6\,L^r_{2})
\nonumber\\&&
       + {A}^{V}(m_\pi^2) \, \Big(  - 15/32\,\frac{1}{16\pi^2}\,m_K^2 + 3/16\,\frac{1}{16\pi^2}\,m_\pi^2 - 3/64\,\overline{A}(m_\pi^2) +
         9/32\,\overline{A}(m_K^2)
\nonumber\\&&
 + 9/64\,\overline{A}(m_\eta^2) + 3/16\,\overline{B}^0(m_\eta^2)\,m_\pi^2 \Big)
\nonumber\\&&
       + {A}^{V}(m_\pi^2)^2 \, \Big(  - 15/128 \Big)
       + {A}^{V}(m_\pi^2)\,{A}^{V}(m_K^2) \, \Big( 3/32 \Big)
       + {A}^{V}(m_\pi^2)\,{A}^{V}(m_\eta^2) \, \Big( 9/64 \Big)
\nonumber\\&&
       + {A}^{V}(m_\pi^2)\,{B}^{0V}(m_\pi^2) \, \Big(  - 3/16\,m_\pi^2 \Big)
       + {A}^{V}(m_\pi^2)\,{B}^{0V}(m_\eta^2) \, \Big( 3/16\,m_\pi^2 \Big)
\nonumber\\&&
       + {A}^{V}(m_K^2) \, \Big(  - 9/16\,\frac{1}{16\pi^2}\,m_K^2 - 3/8\,\frac{1}{16\pi^2}\,m_\pi^2 + 27/32\,\overline{A}(m_\pi^2) + 9/
         16\,\overline{A}(m_K^2)
\nonumber\\&&
 - 9/32\,\overline{A}(m_\eta^2) - 1/2\,\overline{B}^0(m_\eta^2)\,m_K^2 \Big)
\nonumber\\&&
       + {A}^{V}(m_K^2)^2 \, \Big( 3/32 \Big)
       + {A}^{V}(m_K^2)\,{A}^{V}(m_\eta^2) \, \Big(  - 9/32 \Big)
       + {A}^{V}(m_K^2)\,{B}^{0V}(m_\eta^2) \, \Big(  - 1/2\,m_K^2 \Big)
\nonumber\\&&
       + {A}^{V}(m_\eta^2) \, \Big(  - 3/32\,\frac{1}{16\pi^2}\,m_K^2 - 3/16\,\frac{1}{16\pi^2}\,m_\pi^2 + 37/64\,\overline{A}(m_\pi^2) -
         11/32\,\overline{A}(m_K^2)
\nonumber\\&&
 + 9/64\,\overline{A}(m_\eta^2) + 1/3\,\overline{B}^0(m_\eta^2)\,m_K^2 - 7/48\,\overline{B}^0(m_\eta^2)\,m_\pi^2 \Big)
\nonumber\\&&
       + {A}^{V}(m_\eta^2)^2 \, \Big( 9/128 \Big)
       + {A}^{V}(m_\eta^2)\,{B}^{0V}(m_\pi^2) \, \Big( 1/16\,m_\pi^2 \Big)
       + {A}^{V}(m_\eta^2)\,{B}^{0V}(m_K^2) \, \Big(  - 1/4\,m_K^2 \Big)
\nonumber\\&&
       + {A}^{V}(m_\eta^2)\,{B}^{0V}(m_\eta^2) \, \Big( 1/3\,m_K^2 - 7/48\,m_\pi^2 \Big)
\nonumber\\&&
       + {H}^V(m_\pi^2,m_\pi^2,m_K^2,m_K^2) \, \Big( 3/64\,m_K^2 - 3/8\,m_\pi^2 \Big)
       + {H}^V(m_\pi^2,m_K^2,m_\eta^2,m_K^2) \, \Big(  - 9/32\,m_K^2 \Big)
\nonumber\\&&
       + {H}^V(m_K^2,m_K^2,m_K^2,m_K^2) \, \Big(  - 3/8\,m_K^2 \Big)
       + {H}^V(m_K^2,m_\eta^2,m_\eta^2,m_K^2) \, \Big(  - 9/64\,m_K^2 \Big)
\nonumber\\&&
       + {H}_{27}^V(m_\pi^2,m_\pi^2,m_K^2,m_K^2) \, \Big( 9/8 \Big)
       + {H}_{27}^V(m_K^2,m_\pi^2,m_\pi^2,m_K^2) \, \Big(  - 9/64 \Big)
\nonumber\\&&
       + {H}_{27}^V(m_K^2,m_\pi^2,m_\eta^2,m_K^2) \, \Big( 27/32 \Big)
       + {H}_{27}^V(m_K^2,m_K^2,m_K^2,m_K^2) \, \Big( 9/8 \Big)
\nonumber\\&&
       + {H}_{27}^V(m_K^2,m_\eta^2,m_\eta^2,m_K^2) \, \Big( 27/64 \Big)
       + {H}^{V\prime}(m_\pi^2,m_\pi^2,m_K^2,m_K^2) \, \Big(  - 15/64\,m_K^4 + 3/8\,m_\pi^2\,m_K^2 \Big)
\nonumber\\&&
       + {H}^{V\prime}(m_\pi^2,m_K^2,m_\eta^2,m_K^2) \, \Big( 13/32\,m_K^4 \Big)
       + {H}^{V\prime}(m_K^2,m_K^2,m_K^2,m_K^2) \, \Big( 3/8\,m_K^4 \Big)
\nonumber\\&&
       + {H}^{V\prime}(m_K^2,m_\eta^2,m_\eta^2,m_K^2) \, \Big( 181/576\,m_K^4 \Big)
       + {H}_1^{V\prime}(m_K^2,m_\pi^2,m_\pi^2,m_K^2) \, \Big( 3/8\,m_K^4 \Big)
\nonumber\\&&
       + {H}_1^{V\prime}(m_K^2,m_\pi^2,m_\eta^2,m_K^2) \, \Big(  - 3/4\,m_K^4 \Big)
       + {H}_1^{V\prime}(m_K^2,m_\eta^2,m_\eta^2,m_K^2) \, \Big(  - 5/8\,m_K^4 \Big)
\nonumber\\&&
       + {H}_{21}^{V\prime}(m_\pi^2,m_\pi^2,m_K^2,m_K^2) \, \Big( 9/8\,m_K^4 \Big)
       + {H}_{21}^{V\prime}(m_K^2,m_\pi^2,m_\pi^2,m_K^2) \, \Big(  - 9/64\,m_K^4 \Big)
\nonumber\\&&
       + {H}_{21}^{V\prime}(m_K^2,m_\pi^2,m_\eta^2,m_K^2) \, \Big( 27/32\,m_K^4 \Big)
       + {H}_{21}^{V\prime}(m_K^2,m_K^2,m_K^2,m_K^2) \, \Big( 9/8\,m_K^4 \Big)
\nonumber\\&&
       + {H}_{21}^{V\prime}(m_K^2,m_\eta^2,m_\eta^2,m_K^2) \, \Big( 27/64\,m_K^4 \Big)
       + {H}_{27}^{V\prime}(m_\pi^2,m_\pi^2,m_K^2,m_K^2) \, \Big(  - 9/8\,m_K^2 \Big)
\nonumber\\&&
       + {H}_{27}^{V\prime}(m_K^2,m_\pi^2,m_\pi^2,m_K^2) \, \Big( 9/64\,m_K^2 \Big)
       + {H}_{27}^{V\prime}(m_K^2,m_\pi^2,m_\eta^2,m_K^2) \, \Big(  - 27/32\,m_K^2 \Big)
\nonumber\\&&
       + {H}_{27}^{V\prime}(m_K^2,m_K^2,m_K^2,m_K^2) \, \Big(  - 9/8\,m_K^2 \Big)
       + {H}_{27}^{V\prime}(m_K^2,m_\eta^2,m_\eta^2,m_K^2) \, \Big(  - 27/64\,m_K^2 \Big)\,.
\ea

\ba
\lefteqn{F_\pi^3 \Delta^V F_\eta^{(6)} =
        {A}^{V}(m_\pi^2) \, \Big( 2\,L^r_{5}m_\pi^2 + 12\,L^r_{4}m_\pi^2 - 6\,L^r_{3}m_\pi^2 - 6\,L^r_{2}m_\pi^2 - 24
         \,L^r_{1}m_\pi^2 \Big)}
\nonumber\\&&
       + {A}^{V}(m_K^2) \, \Big( 8/3\,L^r_{5}m_K^2 + 4\,L^r_{5}m_\pi^2 + 16\,L^r_{4}m_K^2 - 14\,L^r_{3}m_K^2
          - 8\,L^r_{2}m_K^2 - 32\,L^r_{1}m_K^2 \Big)
\nonumber\\&&
       + {A}^{V}(m_\eta^2) \, \Big( 32/9\,L^r_{5}m_K^2 - 14/9\,L^r_{5}m_\pi^2 + 16/3\,L^r_{4}m_K^2 - 4/3\,
         L^r_{4}m_\pi^2 - 8\,L^r_{3}m_K^2 + 2\,L^r_{3}m_\pi^2
\nonumber\\&&
 - 16\,L^r_{2}m_K^2 + 4\,L^r_{2}m_\pi^2 - 16\,L^r_{1}m_K^2
          + 4\,L^r_{1}m_\pi^2 \Big)
\nonumber\\&&
       + {A}_{23}^{V}(m_\pi^2) \, \Big( 6\,L^r_{3}+ 18\,L^r_{2})
       + {A}_{23}^{V}(m_K^2) \, \Big( 2\,L^r_{3}+ 24\,L^r_{2})
       + {A}_{23}^{V}(m_\eta^2) \, \Big( 6\,L^r_{3}+ 12\,L^r_{2}+ 12\,L^r_{1})
\nonumber\\&&
       + {A}^{V}(m_K^2) \, \Big(  - 3/2\,\frac{1}{16\pi^2}\,m_K^2 - 3/8\,\frac{1}{16\pi^2}\,m_\pi^2 + 3/2\,\overline{A}(m_\pi^2) + 3/4\,
         \overline{A}(m_K^2) \Big)
\nonumber\\&&
       + {A}^{V}(m_\eta^2) \, \Big( 1/2\,\frac{1}{16\pi^2}\,m_K^2 - 1/2\,\overline{A}(m_K^2) \Big)
       + {A}^{V}(m_\eta^2)\,{B}^{0V}(m_K^2) \, \Big(  - 1/2\,m_K^2 \Big)
\nonumber\\&&
       + {H}^V(m_\pi^2,m_K^2,m_K^2,m_\eta^2) \, \Big(  - 9/16\,m_\pi^2 \Big)
       + {H}^V(m_K^2,m_K^2,m_\eta^2,m_\eta^2) \, \Big(  - 3/4\,m_K^2 + 3/16\,m_\pi^2 \Big)
\nonumber\\&&
       + {H}_{27}^V(m_\pi^2,m_K^2,m_K^2,m_\eta^2) \, \Big( 27/16 \Big)
       + {H}_{27}^V(m_\eta^2,m_K^2,m_K^2,m_\eta^2) \, \Big( 27/16 \Big)
\nonumber\\&&
       + {H}^{V\prime}(m_\pi^2,m_\pi^2,m_\eta^2,m_\eta^2) \, \Big( 1/12\,m_\pi^4 \Big)
       + {H}^{V\prime}(m_\pi^2,m_K^2,m_K^2,m_\eta^2) \, \Big( 3/4\,m_\pi^2\,m_K^2 + 1/16\,m_\pi^4 \Big)
\nonumber\\&&
       + {H}^{V\prime}(m_K^2,m_K^2,m_\eta^2,m_\eta^2) \, \Big( 25/9\,m_K^4 - 11/6\,m_\pi^2\,m_K^2 + 5/16\,m_\pi^4
          \Big)
\nonumber\\&&
       + {H}^{V\prime}(m_\eta^2,m_\eta^2,m_\eta^2,m_\eta^2) \, \Big( 64/243\,m_K^4 - 56/243\,m_\pi^2\,m_K^2 + 49/972
         \,m_\pi^4 \Big)
\nonumber\\&&
       + {H}_1^{V\prime}(m_\pi^2,m_K^2,m_K^2,m_\eta^2) \, \Big(  - 2\,m_\pi^2\,m_K^2 + 1/2\,m_\pi^4 \Big)
\nonumber\\&&
       + {H}_1^{V\prime}(m_\eta^2,m_K^2,m_K^2,m_\eta^2) \, \Big(  - 16/3\,m_K^4 + 10/3\,m_\pi^2\,m_K^2 - 1/2\,
         m_\pi^4 \Big)
\nonumber\\&&
       + {H}_{21}^{V\prime}(m_\pi^2,m_K^2,m_K^2,m_\eta^2) \, \Big( 3\,m_K^4 - 3/2\,m_\pi^2\,m_K^2 + 3/16\,m_\pi^4 \Big)
\nonumber\\&&
       + {H}_{21}^{V\prime}(m_\eta^2,m_K^2,m_K^2,m_\eta^2) \, \Big( 3\,m_K^4 - 3/2\,m_\pi^2\,m_K^2 + 3/16\,m_\pi^4 \Big)
\nonumber\\&&
       + {H}_{27}^{V\prime}(m_\pi^2,m_K^2,m_K^2,m_\eta^2) \, \Big(  - 9/4\,m_K^2 + 9/16\,m_\pi^2 \Big)
\nonumber\\&&
       + {H}_{27}^{V\prime}(m_\eta^2,m_K^2,m_K^2,m_\eta^2) \, \Big(  - 9/4\,m_K^2 + 9/16\,m_\pi^2 \Big)\,.
\ea

\end{document}